\documentclass[11pt]{amsart}
\usepackage{mathrsfs}

\usepackage{amscd}
\usepackage{amsmath}
\usepackage{graphicx}
\usepackage{amsfonts}
\usepackage{amssymb}
\begin{document}

\title[Maps preserving strict convex combinations]
{Quantum measurements and maps preserving strict convex combinations
and pure states}

\author{Lihua Yang, Jinchuan Hou}
\address[L. Yang, J. Hou]{Department of Mathematics, Taiyuan University of Technology ,
Taiyuan 030024, P. R.  China} \email{ylhflower@163.com;
jinchuanhou@yahoo.com.cn; houjinchuan@tyut.edu.cn}



\thanks{This work is partially supported by  Natural
Science Foundation of China (11171249, 11271217), and a grant from
International Cooperation Program in Sciences and Technology  of
Shanxi (2011081039).}

\begin{abstract}

In this paper, a characterization of maps between quantum states
that preserve pure states and strict convex combinations is
obtained. Based on this characterization, a structural theorem for
maps between multipartite quantum states that preserve separable
pure states and strict convex combinations is established. Then
these results are applied to characterize injective (local) quantum
measurements and answer some conjectures proposed in
[J.Phys.A:Math.Theor. 45 (2012) 205305].

\end{abstract}
\maketitle

\section{Introduction}

In the theory of quantum mechanics, a state is a positive operator
of trace 1 acting on a complex Hilbert space $H$. Denote by
${\mathcal S}(H)$ and $\mathcal {P}ur(H)$  respectively the set of
all states and the set of all pure states ($i.e.$ rank-1
projections) on $H$. In quantum information theory we deal, in
general, with multipartite systems. The underlying space $H$ of   a
multipartite composite quantum system is a tensor product of
underlying spaces $H_{i}$ of its subsystems, that is $H=H_{1}\otimes
H_{2}\otimes\cdots \otimes H_{n}$. If $n=2$, the system is called a
bipartite system. The definition of multipartite separability was
introduced in \cite{VPJK} as a natural extension of the notion of
separability in bipartite case \cite{W}. Let us denote the set of
all states in an $n$-partite system by ${\mathcal
S}(H_1\otimes\cdots\otimes H_n)$. In the case $\dim H<\infty$, a
state $\rho\in{\mathcal S}(H_1\otimes\cdots\otimes H_n)$ is said to
be (fully) separable if it admits a representation of the form
$$
\rho=\sum_ip_i\rho^{(1)}_i\otimes\cdots\otimes \rho^{(n)}_i,
$$
 where $p_i>0$ with $\sum_ip_i=1$ and $\rho_i^{(k)}\in{\mathcal
S}(H_k)$. Otherwise, $\rho$ is said to be entangled.
 Denote  by respectively ${\mathcal S}_{\rm sep}(H_1\otimes H_2\otimes\cdots\otimes H_n)$ and  ${\mathcal Pur}_{\rm sep}(H_1\otimes H_2\otimes\cdots\otimes H_n)$ the set of all
separable states and the set of all separable pure states on
$H_1\otimes H_2\otimes\cdots\otimes H_n$. It is obvious that
$$\begin{array}{rl} & {\mathcal Pur}_{\rm sep}(H_1\otimes H_2\otimes\cdots\otimes
H_n)=
 {\mathcal Pur}(H_1)\otimes{\mathcal
Pur}(H_2)\otimes\cdots\otimes{\mathcal Pur}(H_n)\\ = &\{P_1\otimes
P_2\otimes \cdots\otimes P_n : P_i\in{\mathcal Pur}(H_i),
i=1,2,\ldots, n\}.\end{array}$$

The theory of maps on the set of states plays an important role in
quantum computation and quantum information science. It is important
to understand, characterize, and construct different classes of maps
on states. For instance, all quantum channels and quantum operations
are completely positive linear maps; in quantum error correction,
one has to construct the recovery map for a given channel; to study
the entanglement of states, one constructs NCP (non completely
positive) positive maps and entanglement witnesses. Many researchers
pay their attention to the problem of characterizing the maps on the
states; ref. \cite{AS,BZ,FLPS,H1}.

The present paper is motivated by some conjectures proposed in
\cite{HL} where the bijective maps $\Phi :{\mathcal S}_{\rm
sep}(H_1\otimes H_2)\rightarrow{\mathcal S}_{\rm sep}(H_1\otimes
H_2)$ that preserve strict convex combinations were studied.

 Recall that a map $\phi$ between convex
sets is said to be (strict) convex combination preserving if, for
any $\rho,\sigma\in {\mathcal S}(H)$ and $t\in[0,1]$ ($t\in(0,1)$),
there is some $s$ with $0\leq s\leq 1$ ($0<s<1$) such that
$\phi(t\rho+(1-t)\sigma)=s\phi(\rho)+(1-s)\phi(\sigma)$. It is
obvious that $\phi$   preserves (strict) convex combination if and
only if $\phi([\rho,\sigma])\subseteq [\phi(\rho),\phi(\sigma)]$
($\phi((\rho,\sigma))\subseteq (\phi(\rho),\phi(\sigma))$), where
$[A,B]$ stands for the closed (open) line segment joint $A$ and $B$,
that is, $[A,B]=\{tA+(1-t)B: 0\leq t\leq 1\}$ ($(A,B)=[A,B]\setminus
\{A,B\}$, here we define $(\rho,\rho)=\{\rho\}$). The maps preserve
strict convex combinations  are closely related to quantum
measurements. In quantum mechanics a fine-grained quantum
measurement is described by a collection $\{M_m\}$ of measurement
operators acting on the Hilbert space $H$ corresponding to the
system satisfying $\sum_mM_m^*M_m=I$; ref. for example, \cite{G}.
Let $M_j$ be a measurement operator. If the state of the quantum
system is $\rho\in{\mathcal S}(H)$ before the measurement, then the
state after the measurement is $\frac{M_j\rho M_j^*}{{\rm
Tr}(M_j\rho M_j^*)}$ whenever $M_j\rho M_j^*\not=0$. If we fix an
$M_j=M$ is fixed, we get a measurement  map $\phi $ defined by $\phi
(\rho)=\frac{M \rho M ^*}{{\rm Tr}(M \rho M ^*)}$ from the convex
subset ${\mathcal S}_{M }(H)=\{ \rho : M \rho M ^*\not=0\}$ of the
(convex) set ${\mathcal S}(H)$  into ${\mathcal S}(H)$, which is a
map that preserves the strict convex combinations and sends  pure
states to pure states. If $M $ is invertible (injective), then $\phi
:{\mathcal S}(H)\rightarrow{\mathcal S}(H)$ is bijective (injective)
and will be called an invertible (injective) measurement map.

The problem of characterizing the strict convex combination maps
between quantum states was firstly attacked by \cite{HHL}. It is
shown in \cite{HHL} that a bijective map $\phi:{\mathcal S}(H)
\rightarrow {\mathcal S}(H)$, $\dim H\geq 2$, is (strict) convex
combination preserving if and only if $\phi$ is an invertible
quantum measurement map or the composition of transpose and an
invertible quantum measurement map. Note that $\phi:{\mathcal
S}(H)\rightarrow{\mathcal S}(H)$ is bijective and (strict) convex
combination preserving imply that $\phi$ preserves pure states in
both directions, that is, $\phi({\mathcal Pur}(H))={\mathcal
Pur}(H)$. Let $\psi: {\mathcal S}_{\rm sep}(H_1\otimes
H_2)\rightarrow{\mathcal S}_{\rm sep}(H_1\otimes H_2)$ be a
bijective map. In \cite{HL}, based on the work of \cite{HHL}, it is
shown that, if $\Phi$ is (strict) convex combination preserving and
if
$$\begin{array}{rl} & \psi(P_1\otimes
P_2)\ \mbox{ is a product state for any}\ P_i\in{\mathcal S}(H_i)\
\mbox{\rm with} \\ & {\rm rank}P_i=1 \ \mbox{ and } {\rm rank}P_j=2,
\ \ (1\leq i\not=j\leq 2),\end{array} \eqno(1.1)$$
 then $\psi$ is a
composition of an invertible local quantum measurement (i.e., the
map of the form $\rho \mapsto\frac{(S\otimes T)\rho (S\otimes
T)^*}{{\rm Tr}((S\otimes T)\rho (S\otimes T)^*)}$ with $S,T$
invertible) and some of the following maps: the transpose, the
partial transpose and the swap.  In the sequel we say a  map is
essentially a (local) quantum measurement if it is a (local) quantum
measurement or a composition of a (local) quantum measurement with
any one of the following maps: the transpose, the partial transpose,
and the swap.

It is conjectured in \cite{HL} that if a bijective map $\psi:
{\mathcal S}_{\rm sep}(H_1\otimes H_2)\rightarrow{\mathcal S}_{\rm
sep}(H_1\otimes H_2)$   is (strict) convex combination preserving,
then it sends in fact product states to product states, and thus the
additional assumption Eq.(1.1) in the result just mentioned above is
superfluous.

The purpose of the present paper is  to answer these conjectures
affirmatively for finite dimensional systems. To do this, we need
first to characterize general maps between states that preserve the
strict convex combinations and send pure states to pure states. Let
$H$ and $K$ be   complex Hilbert spaces of dimension $\geq 2$ with
$\dim H<\infty$. We show in Section 2 that a map
 $\psi:{\mathcal S}(H)\rightarrow{\mathcal S}(K)$   preserves strict convex combinations and pure
 states if and only if it
has one of the following three forms: (1) $\psi$ is contractive to a
pure state, i.e, there exists a pure state $Q\in{\mathcal Pur}(K)$
such that $\psi(\rho)=Q$ for all $\rho\in{\mathcal S}(H)$; (2) there
exist distinct pure states $Q_i\in{\mathcal Pur}(K)$, $i=1,2$ such
that $\psi({\mathcal Pur}(H))=\{Q_1,Q_2\}$, and $\psi({\mathcal
S}(H))\subseteq [Q_1,Q_2]$; (3)   $\dim H \leq \dim K$ and there
exists an injective operator $M\in \mathcal{B}(H,K)$  such that
$\psi(\rho)=\frac{M\rho M^*}{{\rm Tr}(M\rho M^*)}$ for all
$\rho\in{\mathcal S}(H )$, or $\psi(\rho)= \frac{M\rho^tM^*}{{\rm
Tr}(M\rho^tM^*)}$ for all $\rho\in{\mathcal S}(H )$, where $A^t$ is
the transpose of $A$ with respect to an arbitrarily fixed
orthonormal basis of $H$, that is, $\psi$ is essentially an
injective quantum measurement (see Theorem 2.5).  Note that, by our
result, if $\psi$ is strict convex combination preserving and pure
state preserving and if $\psi$ is not continuous, then $\psi$ must
have the form (2), and an example of such map is given (Remark 2.7).
Based on the results obtained in Section 2, we are able to give a
structure theorem of maps $\psi:{\mathcal S}_{\rm sep}(H_1\otimes
H_2)\rightarrow{\mathcal S}_{\rm sep}(K_1\otimes K_2)$ that preserve
strict convex combinations and separable pure states in Section 3,
where $2\leq \dim H_i<\infty$ and $\dim K_i\geq 2$, $i=1,2$. We show
that such maps can have ten possible forms (Theorem 3.2).
Consequently, if the range of $\psi$ is non-collinear or a
singleton,  then $\psi$ sends product states to product states
(Corollary 3.3); and moreover, if the range of $\psi$ also contains
a state $\sigma$ so that  both of its reductions ${\rm
Tr}_1(\sigma)$ and ${\rm Tr}_2(\sigma)$ have rank $\geq 2$, then
$\psi$ is essentially an injective local quantum measurement
(Corollary 3.4). These results particularly answer the conjectures
in \cite{HL}  mentioned above. Section 4 is a brief discussion of
the same topic for multipartite systems. The similar structure
theorem is valid for maps $\psi:{\mathcal S}_{\rm sep}(H_1\otimes
H_2\otimes\cdots\otimes H_n)\rightarrow{\mathcal S}_{\rm
sep}(K_1\otimes K_2\otimes\cdots\otimes K_n)$ that preserve strict
convex combinations and separable pure states, but with more
complicated expressions. Particularly, if the range of $\psi$ is
non-collinear or a singleton, then $\psi$ sends product states to
product states; and moreover, if the range of $\psi$ contains a
state $\sigma$ so that each reduction ${\rm Tr}^i(\sigma)$   has
rank $\geq 2$, then $\psi$ is essentially a local injective quantum
measurement (Theorem 4.1, Corollary 4.2). Section 5 is a short
conclusion.

\section{Maps  preserving pure states and strict convex combinations}

In this section we   characterize the maps between the convex sets
of quantum states that send pure states to pure states and preserve
the strict convex combinations.

We start by giving a simple lemma which is easily checked.

\textbf{Lemma 2.1.} {\it Let $\{Q_1,Q_2,\cdots,Q_r\}$ be a linearly
independent set of rank one projections acting on a Hilbert space
$H$. If $\sum_{i=1}^r t_iQ_i=P_0$ is a projection for some
$t_i>0$,$i=1,2,\ldots, r$, then $\{Q_i\}_{i=1}^r$ are orthogonal and
$t_i=1$ for all $i$. }

\if {\bf Proof.} There are unit vectors $x_i\in H$ such that
$Q_i=x_i\otimes x_i$. If $\sum_{i=1}^r t_iQ_i=P_0$ is a projection
for some $t_i>0$,  then $P^{2}_{0}=P_0\geq 0$, and thus $(\sum
_{i=1}^r t_iQ_i)^2=\sum_{i=1}^r {t_i}^2 x_i\otimes x_i+\sum_{i\ne j}
t_it_j\langle x_j,x_i\rangle x_i\otimes x_j=\sum_{i=1}^r
t_ix_i\otimes x_i.$  This implies that
$$\sum_{i\ne j}t_it_j\langle x_j,x_i\rangle x_i\otimes
x_j=\sum_{i=1}^r\left( t_i-t_i^2\right) x_i\otimes x_i  \eqno
{(2.1)}$$  Since $\{ x_i\} _{i=1}^r $ is a linearly independent set,
there exist $\{y_j\}_{j=1}^r\subset H$, such that $\langle
y_j,x_i\rangle =\delta_{ij}$.  Letting the two sides of Eq.(2.1) act
at $y_j$ $(1\le j\le r)$ respectively, we get $\sum_{i\ne
j}t_it_j\langle x_j,x_i\rangle x_i=(t_j-t_j^2)x_j$. It follows  that
$t_j-t_j^2=0$ and $t_it_j\langle x_j,x_i\rangle =0$. Thus $t_j=1$
for all $j$ and $\langle x_j,x_i\rangle =0$  for any $i\not=j$, that
is, $\{Q_i\}_{i=1}^r$ is an orthogonal set of rank one projections.
\hfill$\Box$ \fi

 Let ${\bf H}_m$ be the real linear space of all
$m\times m$ Hermitian matrices and let ${\mathcal P}_m$ be the set
of all rank-1 $m\times m$ projection matrices. The next lemma comes
from \cite{FLPS} which can be viewed as a characterization of linear
preservers of pure states. Also, ref. \cite{HQ} for infinite
dimensional case.

{\bf Lemma 2.2.} {Suppose $\phi :{\bf H}_m\rightarrow {\bf H}_n$ is
a linear map  satisfying $\phi({\mathcal P}_m)\subseteq{\mathcal
P}_n$. Then one of the following holds:}

(i) {\it There is $Q\in{\mathcal P}_n$ such that $\phi (A)={\rm
Tr}(A)Q$ for all $A\in{\bf H}_m$.}

(ii) {\it $m\leq n$ and there is a $U\in M_{n\times m}$ with
$U^*U=I_m$ such that $\phi(A)=UAU^*$ for all $A\in{\bf H}_m$, or
$\phi(A)=UA^tU^*$ for all $A\in{\bf H}_m$.}

The following lemma is the main result in \cite{Z}, which gives a
characterization of strict convex combination preserving maps in
terms of linear ones.

{\bf Lemma 2.3.} {\it Let $X$ and $Y$ be real linear spaces and
$D\subseteq X$ a nonempty convex subset. Assume that
$\phi:D\rightarrow Y$ is a strict convex combination preserving map
such that $\phi(D)$ is non-collinear (i.e., $\phi(D)$ contains a
nondegenerate triangle). Then, there exist a linear transformation
$A: X\rightarrow Y$, a linear functional $f:X\rightarrow{\mathbb
R}$, a vector $y_0\in Y$, and a scalar $b\in{\mathbb R}$ such that
$$ f(x)+b>0\quad \mbox{for all } \quad x\in D$$ and
$$\phi(x)=\frac{Ax+y_0}{f(x)+b}\quad \mbox{for all } \quad x\in D.$$ }

By using of Lemma 2.1 and Lemma 2.3 we can prove the following
lemma, which is also crucial for proving our main result.

{\bf Lemma 2.4.} {\it Let $H$ be a  complex Hilbert space with
$2\leq \dim H=r<\infty$ and   $\phi:{\mathcal S}(H)\rightarrow
{\mathcal S}(H)$ be a map preserving pure state and strict convex
combinations. If $\phi(\frac 1r I)=\frac 1r I$ and ${\rm ran}\phi$
is non-collinear, then $\phi$ is affine.}

{\bf Proof.} As $\phi$ preserves pure states and strict convex
combinations, by Lemma 2.1, $\phi(\frac 1r I)=\frac 1r I$ implies
that $\phi$ maps orthogonal pure states to orthogonal pure states.
Also, by Lemma 2.3, $\phi$ is strict convex combination preserving
and ${\rm ran}\phi$ is non-collinear together imply that $\phi$ has
the form $\phi(\rho)=\frac{\Gamma (\rho)+D}{f(\rho)+d}$ for any
$\rho\in {\mathcal S}(H)$, where $\Gamma:\mathcal
{B}_{sa}(H)\rightarrow \mathcal {B}_{sa}(H)$ is a linear
transformation, $f:\mathcal {B}_{\rm sa}(H)\rightarrow \mathbb{R}$
is a linear functional, $D\in \mathcal {B}_{\rm sa}(H)$ and $d\in
\mathbb{R}$ with $f(\rho)+d>0$ for all $\rho\in {\mathcal S}(H)$,
${\mathcal B}_{\rm sa}(H)$ is the real linear space of all
self-adjoint operators in ${\mathcal B}(H)$. Since $\dim H<\infty$,
$\Gamma$ and $f$ are continuous. It follows that $\phi$ is
continuous. To prove the lemma, we consider two cases of $\dim H>2$
and $\dim H=2$ respectively.

{\bf Case 1.} $\dim H>2$.

We will show that $f$ is a constant on ${\mathcal S}(H)$, that is,
there is a real number $a$ such that $f(\rho)=a$ for all $\rho\in
{\mathcal S}(H)$.

For any normalized orthogonal basis $\{e_i\}_{i=1}^r$ of $H$, let
$P_i=e_i\otimes e_i$. We first claim that $f(e_i\otimes
e_i)=f(e_j\otimes e_j)$ for any $i$ and $j$. Since $\phi$ preserves
pure states, there is a pure state $Q_i=x_i\otimes x_i$ such that
$$x_i\otimes x_i=Q_i=\phi(P_i)=\frac{\Gamma(e_i\otimes
e_i)+D}{f(e_i\otimes e_i)+d}.$$ So $$\Gamma(e_i\otimes
e_i)+D=(f(e_i\otimes e_i)+d)(x_i\otimes x_i).$$ As $\phi(\frac
Ir)=\frac Ir$ and $\frac Ir =\frac 1r \sum_{i=1}^r e_i\otimes e_i$,
we have $$\frac Ir=\phi(\frac 1r\sum_{i=1}^r e_i\otimes
e_i)=\frac{\Gamma (\sum_{i=1}^r\frac 1r e_i\otimes e_i
)+D}{f(\sum_{i=1}^r\frac 1r e_i\otimes e_i)+d}=\frac{\sum_{i=1}^r
\frac 1r \Gamma (e_i\otimes e_i)+r \frac 1r D}{\sum_{i=1}^r \frac 1r
f(e_i\otimes e_i)+r \frac 1r d},$$ that is, $$\frac Ir =\frac{\frac
1r (\sum_{i=1}^r(\Gamma(e_i\otimes e_i)+D))}{\frac 1r(\sum_{i=1}^r
(f( e_i\otimes e_i)+d))}=\frac {\sum _{i=1}^r(\Gamma (e_i\otimes
e_i)+D)}{\sum_{i=1}^r(f(e_i\otimes e_i)+d)}.$$ \if On the other
hand, as $\phi$ sends orthogonal pure states to orthogonal pure
states,  we have
$$\frac Ir=\frac 1r \sum_{i=1}^r \phi(e_i\otimes
e_i)=\frac 1r \sum_{i=1}^r \frac {\Gamma (e_i\otimes
e_i)+D}{f(e_i\otimes e_i)+d}.$$ Thus we get
$$I=\sum_{i=1}^r \frac {\Gamma(e_i\otimes e_i)+D}{f(e_i\otimes
e_i)+d}.\eqno(2.3)$$\fi Let $A_i=\Gamma(e_i\otimes e_i)+D$ and
$a_i=f(e_i\otimes e_i)+d$. Then $A_i=a_iQ_i$ and the above equation
becomes to
$$I=r(\frac {A_1+A_2+\cdots+A_r}{a_1+a_2+\cdots+a_r})=r(\frac {a_1Q_1+a_2Q_2+\cdots+a_rQ_r}{a_1+a_2+\cdots+a_r}).$$
Applying Lemma 2.1,  we see that $\frac{ra_i}{a_1+a_2+\cdots+a_r}=1$
for each $i=1,2,\ldots,r$ and hence
$$a_1=a_2=\cdots=a_r=\frac{a_1+a_2+\cdots+a_r}{r}.$$ This implies
that there is some scalar $a$ such that $f(e_i\otimes e_i)=a$ holds
for all $i$. Now for arbitrary unit vectors $x$, $y\in H$, as $\dim
H>2$, there is a unit vector $z\in H$ such that $z\in [x, y]^\bot$.
It follows from what proved above that $f(x\otimes x)=f(z\otimes
z)=f(y\otimes y)$. So $f(x\otimes x)=a$ for all unit vectors $x\in
H$. Since each state is a convex combination of pure states, by the
linearity of $f$, we get that $f(\rho)=a$ holds for every state
$\rho$. Therefore, we have
$$\phi(\rho)=\frac{\Gamma(\rho)+D}{a+d}$$ holds for all
$\rho$. Then by the linearity of $\Gamma$, it is clear that $\phi$
is affine, i.e., for any states $\rho$, $\sigma$ and scalar
$\lambda$ with $0\leq \lambda \leq 1$,
$\phi(\lambda\rho+(1-\lambda)\sigma)=\lambda\phi(\rho)+(1-\lambda)\phi(\sigma)$.

{\bf Case 2.} $\dim H=2$.

By fixing an orthonormal basis of $H$ we may identify  $\mathcal
{S}(H)$ with $\mathcal {S}_2$, the convex set of $2\times 2$
positive matrices with the trace 1. Then  $\phi:\mathcal
{S}_2\rightarrow \mathcal {S}_2$ is a map preserving pure states and
strict convex combinations satisfying $\phi(\frac 12 I_2)=\frac 12
I_2$. Let us identify $\mathcal {S}_2$ with the Bloch ball
representation $(\mathbb{R}^3)_1=\{(x, y, z)^t\in
\mathbb{R}^3:x^2+y^2+z^2\leq 1\}$ by the following way. Let
$\pi:(\mathbb{R}^3)_1\rightarrow \mathcal {S}_2$ be the map defined
by $$(x, y, z)^t\longmapsto \frac 12 I_2+\frac 12 \left(
\begin{array}{cc}
z   &x-iy\\
x+iy&-z
\end{array} \right).$$
$\pi$ is a bijective affine isomorphism. Note that $v=(x, y, z)^t$
satisfies $x^2+y^2+z^2=1$ if and only if the corresponding matrix
$\pi(v)$ is a pure state, and $0=(0, 0, 0)^t$ if and only if the
corresponding matrix is $\pi(0)=\frac 12 I$. The map $\phi:\mathcal
{S}_2\rightarrow \mathcal {S}_2$ induces a map
$\hat{\phi}:(\mathbb{R}^3)_1\rightarrow (\mathbb{R}^3)_1$ by the
following equation $$\phi(\rho)=\pi(\hat{\phi}(\pi^{-1}(\rho))).$$
Since $\phi$ is pure state and strict convex combination preserving
and continuous,  and $\pi$ is an affine isomorphism, it is easily
checked that   the map $\hat{\phi}$ is strict convex combination
preserving and   maps the surface of $(\mathbb{R}^3)_1$ into the
surface of $(\mathbb{R}^3)_1$. Since $\phi(\frac 12I)=\frac 12 I$,
we have that $\hat{\phi}((0, 0, 0)^t)=(0, 0, 0)^t$. It is also clear
that the range of $\hat{\phi}$ is non-collinear.

Now applying   Lemma 2.3 to $\hat{\phi}$, there exists a linear
transformation $L:\mathbb{R}^3\rightarrow \mathbb{R}^3$, a linear
functional $f:\mathbb{R}^3\rightarrow \mathbb{R}$, a vector $u_0\in
\mathbb{R}^3$ and a scalar $s\in \mathbb{R}$ such that $f((x, y,
z)^t)+s>0$ and $$\hat{\phi}((x, y, z)^t)=\frac{L((x, y,
z)^t)+u_0}{f((x, y, z)^T)+s}$$ for each $(x, y, z)^T\in
(\mathbb{R}^3)_1$. Since $\hat{\phi}((0, 0, 0)^T)=(0, 0, 0)^t$, we
have $u_0=0$ and $s>0$. Furthermore, the linearity of $f$ implies
that there are real scalars $r_1$, $r_2$, $r_3$ such that $f((x, y,
z)^t)=r_1x+r_2y+r_3z$. We claim that $r_1=r_2=r_3=0$ and hence
$f=0$. If not, then there is a vector $(x_0, y_0, z_0)^T$ satisfying
$x_0^2+y_0^2+z_0^2=1$ such that $f((x_0, y_0,
z_0)^T)=r_1x_0+r_2y_0+r_3z_0\neq 0$. It follows that
$$1=\|\hat{\phi}((x_0, y_0, z_0)^t)\|=\|\frac{L((x_0, y_0,
z_0)^t)}{r_1x_0+r_2y_0+r_3z_0+s}\|,$$ and thus $$\|L((x_0, y_0,
z_0)^t)\|=r_1x_0+r_2y_0+r_3z_0+s.$$ Similarly $$\|L((-x_0, -y_0,
-z_0)^t)\|=-r_1x_0-r_2y_0-r_3z_0+s.$$ By the linearity of $L$ we
have $r_1x_0+r_2y_0+r_3z_0+s=-r_1x_0-r_2y_0-r_3z_0+s$. Hence
$r_1x_0+r_2y_0+r_3z_0=0$, a contradiction. So, we have $f=0$, and
thus $\hat{\phi}=\frac {L}{s}$ is affine. Now it is clear that
$\phi$ is affine as $\pi$ is an affine isomorphism. \hfill$\Box$

The following is the main result of this section.

\textbf{Theorem 2.5.} {\it Let $H,K$ be complex Hilbert spaces with
$2\leq\dim H<\infty$ and ${\mathcal S}(H), {\mathcal S}(K)$ the
convex sets of all states on $H,K$, respectively. Let
$\psi:{\mathcal S}(H) \rightarrow {\mathcal S}(K)$ be a map. Then
$\psi$ preserves pure states and strict convex combinations (that
is, $\psi({\mathcal Pur}(H))\subseteq{\mathcal Pur}(K)$ and $\psi
((\rho,\sigma))\subseteq (\psi(\rho),\psi(\sigma))$ for any $\rho,
\sigma\in{\mathcal S}(H)$) if and only if one of the following
holds:}\vspace{2mm}

(1) {\it  There exists $\sigma_0\in {\mathcal Pur}(K)$ such that
$\psi (\rho)=\sigma_0$ for all $\rho\in{\mathcal S}(H)$. }
\vspace{3mm}

(2) {\it There exist distinct pure states $Q_i\in{\mathcal Pur}(K)$,
$i=1,2$ such that $\psi({\mathcal Pur}(H))=\{Q_1,Q_2\}$, and a map
$h:{\mathcal S}(H)\rightarrow [0,1]$ such that, for any
$\rho_1,\rho_2\in{\mathcal S}(H)$ and any $t\in(0,1)$,
$h(t\rho_1+(1-t)\rho_2)=sh(\rho_1)+(1-s)h(\rho_2)$ for some $s\in
(0,1)$,  and $\psi(\rho )=h(\rho)Q_1+(1-h(\rho))Q_2$ for all
$\rho\in{\mathcal S}(H)$.}

(3) {\it $\dim H\leq \dim K$ and there exists an injective operator
$M\in \mathcal{B}(H,K)$  such that $\psi(\rho)=\frac{M\rho M^*}{{\rm
Tr}(M\rho M^*)}$ for all $\rho\in{\mathcal S}(H)$, or $\psi(\rho)=
\frac{M\rho^tM^*}{{\rm Tr}(M\rho^tM^*)}$ for all $\rho\in{\mathcal
S}(H)$, where $A^t$ is the transpose of $A$ with respect to an
arbitrarily fixed orthonormal basis of $H$. }

We remark here that the form (3) can be restated as:

(3$^\prime$) {\it $\dim H\leq \dim K$ and there exists an injective
 linear or conjugate linear operator $M: H\rightarrow K$ such that
$\psi(\rho)=\frac{M\rho M^*}{{\rm Tr}(M\rho M^*)}$ for all $\rho\in
{\mathcal S}(H)$.}\\ This statement is more convenient some times.

The following corollary is immediate, which essentially gives a
characterization of injective quantum measurement maps or the
transpose of an injective quantum measurement. We say that a map
$\psi$ is open line segment preserving if
$\psi((\rho,\sigma))=(\psi(\rho),\psi(\sigma))$ for any $\rho,\sigma
$.

\textbf{Corollary 2.6.} {\it Let $H,K$ be complex Hilbert spaces
with $2\leq\dim H<\infty$ and $\psi:{\mathcal S}(H) \rightarrow
{\mathcal S}(K)$ be a map. Then the following statements are
equivalent.}

(1) {\it $\psi$ is strict convex combination preserving with  $\psi
({\mathcal Pur}(H))\subseteq {\mathcal Pur}(K)$ and  non-collinear
range.}

(2) {\it $\psi$ is open line segment preserving   with $\psi
({\mathcal Pur}(H))\subseteq {\mathcal Pur}(K)$ and non-collinear
range. }

(3) {\it  $\dim H\leq \dim K$ and there exists an injective operator
$M\in \mathcal{B}(H,K)$  such that $\psi(\rho)=\frac{M\rho M^*}{{\rm
Tr}(M\rho M^*)}$ for all $\rho\in{\mathcal S}(H)$, or $\psi(\rho)=
\frac{M\rho^tM^*}{{\rm Tr}(M\rho^tM^*)}$ for all $\rho\in{\mathcal
S}(H)$, where $A^t$ is the transpose of $A$ with respect to an
arbitrarily fixed orthonormal basis of $H$.}\vspace{2mm}

Particularly, if $\psi$ is bijective, then, by \cite[Lemma
2.1]{HHL}, we have $\psi (\mathcal {P}ur(H))={\mathcal Pur}(K)$.
Also the surjectivity of $\psi$ implies the surjectivity of $M$.
Thus the above corollary is a generalization of the main result   in
\cite{HHL} for finite dimensional case.

{\bf Remark 2.7.} If the map $\psi$ has the form  (1) or (3) of
Theorem 2.5, then   $\psi$ is continuous. However, if $\psi $ has
the form (2), $\psi$ is not continuous and may have erratic
behavior. For example, Assume $H$ is of dimension 2. Let $Q_1,Q_2$
be two distinct pure states on $K$. Divide ${\mathcal Pur}(H)$ into
two disjoint parts ${\mathcal Pur}(H)={\mathcal P}_1\cup{\mathcal
P}_2$ with the property $P\in{\mathcal P}_1\Leftrightarrow
P^\bot\in{\mathcal P}_2$ and define $\psi(P)=Q_1$ if $P\in{\mathcal
P}_1$; $\psi(P)=Q_2$ if $P\in{\mathcal P}_2$;
$\psi(tP_1+(1-t)P_2)=\frac{1}{2}(Q_1+Q_2)$ if $P_i\in{\mathcal
P}_i$, $i=1,2$, and $t\in(0,1)$, where $Q_1,Q_2$ are any distinct
pure states on $K$. Then, $\psi:{\mathcal S}(H)\rightarrow{\mathcal
S}(K)$ is strict convex combination preserving and $\psi({\mathcal
Pur}(H))\subset{\mathcal Pur}(K)$. $\psi$ has the form (2) in
Theorem 2.5.

 Now let us start to prove the main result of this
section.

{\bf Proof of Theorem 2.5.}

If $\psi$ has the form (1) or (2)  or (3), it is clear that $\psi$
is pure states and strict convex combination preserving. Conversely,
assume that $\psi$ is pure states and strict convex combination
preserving. We will show that $\psi$ has one of the forms stated in
(1), (2) and (3).

Assume $\dim H=m<\infty$.

As $\frac{1}{m}I\in{\mathcal S}(H)$, where $I$ is the identity on
$H$, $\psi (\frac{1}{m} I)$ is positive with trace 1. So $\psi
(\frac{1}{m} I)=\frac {RR^*}{{\rm Tr}(RR^*)}$ for some bounded
linear operator $R $ from $H$ into $K$.

{\bf Claim  1.} $\dim{\rm ran} R\leq m$ and ${\rm ran}
\psi(\rho)\subseteq {\rm ran} R$ holds for all $\rho \in \mathcal
{S}(H)$.

Firstly we will show that ${\rm ran}\psi(P)\subseteq {\rm ran}R$
holds for any pure state $P\in \mathcal {S}(H)$. Let $P=P_1$. There
exist pure states $\{P_2,\cdots,P_m\}$ such that
$\{P_1,P_2,\cdots,P_m\}$ is an orthogonal set  satisfying
$P_1+P_2+\cdots+P_m=I$. Since $\psi$ is strict convex combination
preserving, there are $p_i\in (0,1)$ $(i=1,2,\cdots,m)$ with
$\sum_{i=1}^mp_i=1$ such that $$\frac {RR^*}{{\rm
Tr}(RR^*)}=\psi(\frac1m I)=\psi (\frac1m
\sum_{i=1}^mP_i)=\sum_{i=1}^mp_i\psi(P_i).$$  Note that $\psi(P_i)$s
are rank-1 projections by the assumption. It follows that $\dim {\rm
ran}R\leq \sum_{i=1}^m\dim{\rm ran}\psi(P_i)\leq m$ and, for any
$i$, we have $0\leq\psi(P_i)\leq p_i^{-1}\frac {RR^*}{{\rm
Tr}(RR^*)}$. Hence ${\rm ran}\psi(P_i)\subseteq {\rm ran}R$ for all
$i$. Particularly, ${\rm ran}\psi(P)={\rm ran}\psi(P_1)\subseteq
{\rm ran}R$.

For any $\rho\in{\mathcal S}(H)$, let $\rho=\sum_{i=1}^m t_iP_i$ be
its spectral resolution. As $t_i\geq 0$, $\sum_i t_i=1$ and $\psi$
is strict convex combination preserving, there are $s_i\in[0,1]$
with $\sum_is_i=1$ and $s_i=0$ if $t_i=0$ such that $\psi(\rho)=\sum
_{i=1}^ms_i\psi(P_i)$. Now it is clear that ${\rm ran}
\psi(\rho)\subseteq {\rm ran} R$ because ${\rm
ran}\psi(P_i)\subseteq {\rm ran}R$ for all $i=1,2,\ldots,m$.

As a result, if $R$ is of rank-1, then it is clear that $\psi$ has
the form (1) in   Theorem 2.5.

So, in the sequel, we assume that rank$(R)=r\geq 2$. Thus we can
define a map $\phi:\mathcal {S}(H)\rightarrow \mathcal {S}(H)$ by
$$\phi(\rho)= \frac{R^{[-1]}\psi(\rho)R^{[-1]*}}{{\rm
Tr}(R^{[-1]}\psi(\rho)R^{[-1]*})}, \eqno(2.2)$$ where $R^{[-1]}$ is
the Moore-Penrose generalized inverse of $R$. It is clear that
$\phi(\mathcal {P}ur(H))\subseteq \mathcal {P}ur(H)$ and $\phi$
preserves strict convex combination. Write $\phi(\frac 1m
I_m)=\frac{Q_0}{{\rm Tr}(Q_0)}$, where $Q_0=R^{[-1]}R$. This implies
that $Q_0$ is a projection with ${\rm rank}Q_0=r\leq m$. It follows
that, there exists an orthonormal set $\{e_1, \ldots, e_r\}\subset
H$ such that $\sum_{i=1}^r\phi(P_i)=Q_0$, where $P_i=e_i\otimes
e_i$. Let $H_1={\rm span}\{e_1,\ldots, e_r\}$ and $K_1=Q_0(H)$. Then
$\dim H_1=\dim K_1=r$ and $\widetilde{\phi}=\phi|_{{\mathcal
S}(H_1)}:{\mathcal S}(H_1)\rightarrow{\mathcal S}(K_1)$ is a strict
convex combination preserver sending pure states to pure states. As
$\widetilde{{\phi}}(\frac{1}{r}I_{H_1})=\sum_{i=1}^r
q_i\phi(P_i)=SS^*$ is an invertible state on $K_1$,  it induces a
strict convex combination preserver $\widehat{{\phi}}:{\mathcal
S}(H_1)\rightarrow{\mathcal S}(H_1)$ satisfying
$\widehat{\phi}({\mathcal Pur}(H_1))\subseteq{\mathcal Pur}(H_1)$
and $\widehat{\phi}(\frac{1}{r}I_{H_1})=\frac{1}{r}I_{H_1}$, where
$\widehat{\phi}$ is defined by
$$\widehat{\phi}(\rho)=\frac{S^{-1}\widetilde{\phi}(\rho) {S^{-1}}^*}{{\rm Tr}(S^{-1}\widetilde{\phi}(\rho) {S^{-1}}^*)}. \eqno(2.3)$$ It follows that
$\widehat{\phi}$ maps orthogonal pure states to orthogonal ones.

{\bf Claim 2.} If $r\geq 3$, then there is a unitary operator $U:
H_1\rightarrow H_1$ such that either $\widehat{\phi}(\rho)=U\rho
U^*$ for every $\rho\in{\mathcal S}(H_1)$ or
$\widehat{\phi}(\rho)=U\rho^t U^*$ for every $\rho\in{\mathcal
S}(H_1)$.

If $r\geq 3$, then $\widehat{\phi}({\mathcal S}(H_1))$ contains at
least 3 rank one projections which are orthogonal to each other and
hence non-collinear.    So, by Lemma 2.4,   $\widehat{\phi}$ is
affine, that is
$\widehat{\phi}(t\rho+(1-t)\sigma)=t\widehat{\phi}(\rho)+(1-t)\widehat{\phi}(\sigma)$
holds for any $\rho,\sigma\in{\mathcal S}(H_1)$ and $t\in[0,1]$.
Thus $\widehat{\phi}$ is injective and can be extended to an
injective linear map from ${\mathcal B}_{\rm sa}\cong{\bf H}_r$ into
${\mathcal B}_{\rm sa}\cong{\bf H}_r$.

Now by Lemma 2.2, there is a unitary operator $U: H_1\rightarrow
H_1$ such that either $\widehat{\phi}(\rho)=U\rho U^*$ for every
$\rho\in{\mathcal S}(H_1)$ or $\widehat{\phi}(\rho)=U\rho^t U^*$ for
every $\rho\in{\mathcal S}(H_1)$.

If $P$ and $Q$ are projections and $PQ=0$, we say that $P$ and $Q$
are orthogonal, denoted by $P\bot Q$. $P^\bot$ stands for $I-P$.

{\bf Claim 3.} If $r=2$, then either

(i) there exist ${\mathcal P}_1,{\mathcal P}_2\subseteq {\mathcal
Pur}(H_1)$ satisfying   ${\mathcal P}_1 \cup{\mathcal P}_2={\mathcal
Pur}(H_1)$ and $P, I_2-P$ can not be in the same ${\mathcal P}_i$;
and there exist $Q_1,Q_2\in{\mathcal Pur}(H_1)$ with $Q_1\bot Q_2$
such that $\widehat{\phi}({\mathcal P}_i)=\{Q_i\}$, $i=1,2$, and
$\widehat{\phi}({\mathcal S}(H_1)\subseteq [Q_1,Q_2]$.

(ii) there exists a unitary operator $U: H_1\rightarrow H_1$ such
that either $\widehat{\phi}(\rho)=U\rho U^*$ for every
$\rho\in{\mathcal S}(H_1)$ or $\widehat{\phi}(\rho)=U\rho^t U^*$ for
every $\rho\in{\mathcal S}(H_1)$.

If there exist $Q_1,Q_2,Q_3\in\widehat{\phi}({\mathcal S}(H_1))$
such that they are non-collinear, then Lemma 2.4 is applicable. It
follows that $\widehat{\phi}$  is affine and can be extended to a
linear or conjugate linear map, still denoted by $\widehat{\phi}$
from ${\bf H}_2$ into ${\bf H}_2$, which is unital and rank-1
projection preserving. \if Thus by \cite{H1,H2},  there exist
$M_i\in {\mathcal B}(K_1,K_2)$, $i=1,\ldots, 4$ such that
$\widehat{\phi}(A)=\sum_{i=1}^3 M_iAM_i^* \pm M_4AM_4^*$ for all
$A\in{\bf H}_2$, or $\widehat{\phi}(A)=\sum_{i=1}^3 M_iA^tM_i^* \pm
M_4A^tM_4^*$ for all $A\in{\bf H}_2$. We may assume that $M_1\not=0$
and has the maximal rank amon $M_i$s. With no loss of generality we
suppose that $\widehat{\phi}$ takes the first form. Since
$\widehat{\phi}$ sends rank-1 projections to rank-1 projections, for
any $x\in K_1$, there is some $y_x\in K_2$ such that
$$\sum_{i=1}^3 M_ix\otimes M_ix \pm M_4x\otimes M_4x=y_x\otimes
y_x,\eqno(2.4)$$ which implies that span$\{M_1x,\ldots, M_4x\}$ is a
linear subspace of at most dimension 1. If rank$(M_1)=2$, then
$M_ix=\lambda_{i,x}M_1x$ for any $x$, which forces that $M_i=\lambda
M_1$ for some scalar $\lambda_i$, $i=2,3,4$. Hence there exists an
operator $V:K_1\rightarrow K_2$ such that $\Phi(A)=VAV^*$ for all
$A$. It is clear that $\|Vx\|=\|x\|$. So, $V$ is an isometry, and
$\widehat{\phi}$ take the form (ii) of the claim. If rank$(M_1)=1$,
then there exist $f_i\in K_1$ and $y_i\in K_2$, $i=1,\ldots, 4$ such
that $M_i=y_i\otimes f_i$, the operators of rank at most 1. We may
require that $f_i\not=0$ for each $i$. It follows from (2.4) that
there exists $y_0\in K_2$ such that $\widehat{\phi}(x\otimes x)
=\sum_{i=1}^3|\langle x,f_i\rangle|^2y_i\otimes y_i\pm |\langle
x,f_4\rangle|^2y_4\otimes y_4=y_0\otimes y_0$ for any $x\in K_1$
with $\|x\|=1$. But this is impossible since
$\widehat{\phi}(\frac{1}{2}I_2)=\frac{1}{2}I_2$.\fi  Therefore, use
Lemma 2.2 we see that (ii) is true.

Assume that $\widehat{\phi}({\mathcal S}(H_1)$ is collinear; then
there exist $P_1,P_2\in {\mathcal S}(H_1)$ such that
$\widehat{\phi}( P_i)=Q_i$, $Q_1+Q_2=I_2$ and
$\widehat{\phi}({\mathcal S}(H_1)\subseteq[Q_1,Q_2]$. It entails
that $\widehat{\phi}({\mathcal Pur}(H_1))=\{Q_1,Q_2\}$. Let
${\mathcal P}_i=\widehat{\phi}^{-1}(\{Q_i\})$, $i=1,2$. Then,
${\mathcal P}_1\cap{\mathcal P}_2=\emptyset$ and ${\mathcal
P}_1\cup{\mathcal P}_2={\mathcal Pur}(H_1)$. If $P\in {\mathcal
P}_1$, then
$$\frac{1}{2}\widehat{\phi}(P)+\frac{1}{2}\widehat{\phi}(P^\bot)
=\widehat{\phi}(\frac{1}{2}(P+P^\bot))=\frac{1}{2}I_2.$$ So
$\widehat{\phi}(P^\bot)=Q_2$, and $P^\bot\in{\mathcal P}_2$. Now, it
is clear that  (i) holds.

{\bf Claim 4.} If $r=m\geq 3$, then  $\psi(\rho)=\frac {M\rho
M^*}{{\rm Tr}(M\rho M^*)}$ for all $\rho\in{\mathcal S}(H)$; or
$\psi(\rho)=\frac {M\rho^t M^*}{{\rm Tr}(M\rho^t M^*)}$ for all
 $\rho\in{\mathcal S}(H)$, where $M  :H\rightarrow K$ is an
 injective linear operator. So, $\psi$ has the form (3) of
 Theorem 2.5.

In fact, in this case we have
$\widehat{\phi}=\widetilde{\phi}=\phi$. By Claim 2, there is a
unitary $U: H\rightarrow H$ such that $\phi(\rho)=U\rho U^*$ for all
$\rho$ or $\phi(\rho)=U\rho^t U^*$ for all $\rho$. Let $M= RU$. Then
$M: H\rightarrow K$ is injective and the claim holds.

By use of Claim 4, the following claim is obvious.

{\bf Claim 5.} If $r=m= 2$, then either $\psi$ has the form (3) or
has the form (2) of Theorem 2.5.

{\bf Claim  6.} If $2= r<m$, then $\psi$ has the form (2).

As $r=2$, $\widehat{\phi}$ has two possible forms (i) and (ii)
stated in Claim 3.

If $\widehat{\phi}$ has the form (i), then there exist distinct pure
states $Q_1,Q_2$ on $K$ such that $\psi({\mathcal
Pur}(H))=\{Q_1,Q_2\}$. It is clear that, in this case, we have
$\psi(\rho)\in[Q_1,Q_2]$ for every $\rho\in{\mathcal S}(H)$, that
is, $\psi$ is of the form (2) stated in Theorem 2.5.

We assert that the case (ii) does not occur. If $\widehat{\phi}$
takes form (ii), then there exists an orthogonal set of pure states
$\{P_1=e_1\otimes e_1,\ldots, P_m=e_m\otimes e_m\}$ such that
ran$(\psi)={\mathcal S}(K_1)$, where $K_1={\rm span}\{u_1,u_2\}$
with $u_i\otimes u_i=\psi(e_i\otimes e_i)$, $i=1,2$. Thus $\psi$ is
continuous when restricted on ${\mathcal S}(H_1)$ with $H_1={\rm
span}\{e_1,e_2\}$. Note that $\psi(P_3)=\psi(P_1)$ or $\psi(P_2)$,
say $\psi(P_3)=\psi(P_1)=Q_1$. Let $P(\alpha,\beta)=(\alpha
e_1+\beta e_2)\otimes (\alpha e_1+\beta e_2)\in {\mathcal
Pur}(H_1)$, where $\alpha,\beta\in{\mathbb C}$ satisfying
$|\alpha|^2+|\beta|^2=1$. Then, $P(\alpha,\beta)$ is continuous and
hence $\psi(P(\alpha,\beta))$ is continuous in $\alpha,\beta$. Since
$\{P(\alpha,\beta),P(\bar{\beta},-\bar{\alpha}), P_3,\ldots, P_m\}$
is still a complete orthogonal set of rank-1 projections, one of
them must be $\psi(P_3)=Q_1$ and another be $Q_2$. It follows that
the range of $\psi(P(\alpha,\beta))$ can take at most two distinct
value and thus must be a constant function. However,
$\psi(P(1,0))=\psi(P_1)=Q_1$ and $\psi(P(0,1))=\psi(P_2)=Q_2$, a
contradiction. So, this case can not occur, finishing the proof of
Claim 6.

{\bf Claim  7.} The case of $3\leq r< m$ can not occur.

On the contrary, suppose $m>r\geq 3$. Then by Lemma 2.3, $\phi$ is
continuous. Choose any orthogonal set of rank one projections
$\{P_i=x_i\otimes x_i\}_{i=1}^m$ satisfying $\sum_{i=1}^mP_i=I$.
Then there exists $p_i> 0$ with $\sum_{i=1}^mp_i=1$ such that
$\phi(\frac {1}{m} I)=\phi(\frac {1}{m}
\sum_{i=1}^mP_i)=\sum_{i=1}^mp_i\phi(P_i)=\frac{1}{r}Q_0$. It
follows that $m$ rank-1 projections $\phi(P_1),\cdots,\phi(P_m)$ are
linearly dependent. Without loss of generality, assume
$\{\phi(P_1),\cdots,\phi(P_r)\}$ is linearly independent. Then for
any $j>r$, there exists $i_j\leq r$ such that
$\phi(P_j)=\phi(P_{i_j})$. So there exist $q_i>0$, $(1\leq i\leq
r)$, such that $\phi(\frac 1m I)=\sum_{i=1}^rq_i\phi(P_i)$. By Lemma
2.1, we obtain that $q_i=\frac 1r$ and $\{\phi(P_i)\}_{i=1}^r$ is an
orthogonal set of rank-1 projections. Consequently we obtain that,
for any two orthogonal rank-1 projections $Q_1=y_1\otimes y_1$,
$Q_2=y_2\otimes y_2$ on $H$, either $\phi(Q_1)=\phi(Q_2)$ or
$\phi(Q_1)\bot \phi(Q_2)$. Let $Q'_1=(\alpha y_1+\beta y_2) \otimes
(\alpha y_1+\beta y_2)$, $Q'_2=(\bar{\beta}y_1-\bar{\alpha}y_2)
\otimes (\bar{\beta}y_1-\bar{\alpha}y_2)$, where
$|\alpha|^2+|\beta|^2=1$. Clearly $Q'_1\bot Q'_2$. We assert that:
$$\phi(Q_1)=\phi(Q_2)\Rightarrow\phi(Q'_1)=\phi(Q'_2)\quad\mbox{\rm
and}\quad
\phi(Q_1)\bot\phi(Q_2)\Rightarrow\phi(Q'_1)\bot\phi(Q'_2).\eqno(2.4)$$
To see this, let $f(\alpha,\beta)=\phi(Q'_1)$,
$g(\alpha,\beta)=\phi(Q'_2)$ and let
$h(\alpha,\beta)=\|f(\alpha,\beta)-g(\alpha,\beta)\|$. As $\phi$ is
continuous, we see that $f,g$ and $h$ are continuous in
$\alpha,\beta$. Also note that $h(\alpha,\beta)\in\{0,1\}$  for any
$(\alpha,\beta)$. Hence, if $\phi(Q_1)=\phi(Q_2)$, then $h(1,0)=0$,
which forces $h(\alpha,\beta)\equiv0$ and consequently,
$f(\alpha,\beta)=g(\alpha,\beta)$ for all $\alpha,\beta$; if
$\phi(Q_1)\bot\phi(Q_2)$, then $h(\alpha,\beta)\equiv h(0,1)=1$,
which implies that $f(\alpha,\beta)\bot g(\alpha,\beta)$ for all
$\alpha,\beta$. So the assertion  (2.4) is true.

Now for the chosen orthogonal set $\{P_i\}_{i=1}^m$, as $3\leq r<m$,
by what proved above, we can rearrange the order of
$\{\phi(P_i)\}_{i=1}^m$
 so that $\{\phi(P_1),\ldots, \phi(P_r)\}$ is an
orthogonal set. Then $\phi(P_1)+\cdots +\phi(P_r)=Q_0$,
$\phi(P_{r+j})$ equals to $\phi(P_i)$ for some $i$ with $1\leq i\leq
r$.

Assume that there exist two distinguished projections in
$\{\phi(P_{r+1}),\ldots,\phi(P_{m})\}$, say $\phi(P_{r+1})\not=
\phi(P_{r+2})$. Let $P^\prime _{r+1}=(\alpha x_{r+1}+\beta
x_{r+2})\otimes (\alpha x_{r+1}+\beta x_{r+2})$ and
$P^\prime_{r+2}=(\bar{\beta} x_{r+1}-\bar{\alpha} x_{r+2})\otimes
(\bar{\beta} x_{r+1}-\bar{\alpha} x_{r+2})$  with
$|\alpha|^2+|\beta|^2=1$. By Eq.(2.4), $f(\alpha,\beta) \bot
g(\alpha,\beta)$, where $f(\alpha,\beta)=\phi(P^\prime _{r+1})$ and
$g(\alpha,\beta)=\phi(P^\prime _{r+2})$. Since $\{P_1,\ldots,P_r,
P^\prime _{r+1},P^\prime _{r+2},\ldots,P_m\}$ is still orthogonal,
we see that $f(\alpha,\beta)\in \{\phi(P_1), \ldots, \phi(P_r)\}$.
The continuity of $f$ then implies that
$f(\alpha,\beta)\equiv\phi(P_{i_0})$ for some $1\leq i_0\leq r$.
Similarly, $g(\alpha,\beta)\equiv\phi(P_{i_1})$ for some $1\leq
i_1\leq r$. Since $f$ and $g$ has the same range, we must have
$i_0=i_1$, but this contradicts to $f(1,0)\bot g(1,0)$.

Therefore, we may assume that
$\phi(P_{r+1})=\cdots=\phi(P_m)=\phi(P_1)$. Now  Let $P^\prime
_{1}=(\alpha x_{1}+\beta x_{2})\otimes (\alpha x_{1}+\beta x_{2})$
and $P^\prime_{2}=(\bar{\beta} x_{1}-\bar{\alpha} x_{2})\otimes
(\bar{\beta} x_{1}-\bar{\alpha} x_{2})$  with
$|\alpha|^2+|\beta|^2=1$. Denote $f(\alpha,\beta)=\phi(P^\prime
_{1})$ and $g(\alpha,\beta)=\phi(P^\prime _{2})$. Then by Eq.(2.4)
again we have $f(\alpha,\beta) \bot g(\alpha,\beta)$. Note that
$f(1,0)=\phi(P_1)=\phi(P_{r+1})\in\{f(\alpha,\beta),
g(\alpha,\beta)\}$ for any $(\alpha,\beta)$. This entails
$\{f(\alpha,\beta), g(\alpha,\beta)\}=\{\phi(P_1),\phi(P_2)\}$ for
any $(\alpha,\beta)$. So it follows from the continuity of $f$ and
$g$ that $f(\alpha,\beta)\equiv\phi(P_1)$ and
$g(\alpha,\beta)\equiv\phi(P_2)$, contradicting to the fact that $f$
and $g$ has the same range. So the claim is true.

Combining Claims 1-7, we   see that $\psi$ preserves pure state and
strict convex combinations will imply that $\psi$ takes one of the
form (1), (2) and (3), completing the proof of Theorem 2.5.
\hfill$\Box$

\section{Maps preserving  separable pure states and strict convex combinations: bipartite systems}

This section is devoted to  giving a  structure theorem of  maps
preserve separable pure states and strict convex combinations for
bipartite systems. Using this structure theorem we are able to
answer the conjectures  for finite dimensional case mentioned in the
introduction section that were proposed in \cite{HL} .

The following  simple lemma  may be found in \cite{MN}.

\textbf{Lemma 3.1}  {\it Let $H$ be a complex Hilbert space of any
dimension and $T\in{\mathcal B}(H)$ a finite rank operator. Then
$\frac {1}{{\rm rank}T} \|T\|_{\rm Tr}^2\leq\|T\|_2^2\leq\|T\|_{\rm
Tr}^2$, where $\|T\|_{\rm Tr}$ and $\|T\|_{2}$ are respectively the
trace-norm and the Hilbert-Schmidt norm of $T$. }

Recall that ${\mathcal S}_{\rm sep}(H_1\otimes H_2)$ and $\mathcal
{P}ur(H_1)\otimes \mathcal {P}ur(H_2)$   stand respectively for the
convex set of all separable states and the set of all separable pure
states in bipartite system $H_1\otimes H_2$. Now let us present the
main result of this section.

\textbf{Theorem 3.2}  {\it Let $H_1,H_2,K_1,K_2$ be complex Hilbert
spaces with $2\leq\dim H_i<\infty$, $i=1, 2$. Let $\psi:{\mathcal
S}_{\rm sep}(H_1\otimes H_2)\rightarrow {\mathcal S}_{\rm
sep}(K_1\otimes K_2)$ be a map. If $\psi$ preserves separable pure
states and strict convex combinations, then one of the following
statements holds.}

(1) {\it There exists $R_1\otimes R_2\in \mathcal {P}ur(K_1)\otimes
\mathcal {P}ur(K_2)$ such that $$\psi(A\otimes B)=R_1\otimes R_2$$
for all $A \in {\mathcal S} (H_1) $ and $B \in {\mathcal S} (H_2)
$.}

(2) {\it $\dim H_1\leq \dim K_1$, there exist $R_2\in \mathcal
{P}ur(K_2)$ and injective $M_1\in \mathcal {B}(H_1,K_1)$ such that
$$\psi(A\otimes B)=\frac{M_1AM_1^*}{{\rm Tr}(M_1AM_1^*)}\otimes R_2 \ \mbox{\rm for all } A \in {\mathcal S} (H_1)  \ \mbox{\rm and}\  B \in {\mathcal S} (H_2)
$$ or $$ \psi(A\otimes
B)=\frac{M_1A^tM_1^*}{{\rm Tr}(M_1A^tM_1^*)}\otimes R_2 \ \mbox{\rm
for all } A \in {\mathcal S} (H_1)  \ \mbox{\rm and}\  B \in
{\mathcal S} (H_2).$$ }

(3) {\it $\dim H_2\leq \dim K_2$, there exist $R_1\in \mathcal
{P}ur(K_1)$ and injective $M_2\in \mathcal {B}(H_2,K_2)$ such that
$$\psi(A\otimes B)=R_1\otimes \frac{M_2BM_2^*}{{\rm
Tr}(M_2BM_2^*)} \ \mbox{\rm for all } A \in {\mathcal S} (H_1)  \
\mbox{\rm and}\  B \in {\mathcal S} (H_2)$$ or $$ \psi(A\otimes
B)=R_1\otimes \frac{M_2B^tM_2^*}{{\rm Tr}(M_2B^tM_2^*)} \ \mbox{\rm
for all } A \in {\mathcal S} (H_1)  \ \mbox{\rm and}\  B \in
{\mathcal S} (H_2).$$}

(4) {\it $\dim H_2\leq \dim K_1$, there exist $R_2\in \mathcal
{P}ur(K_2)$ and injective $M_1\in \mathcal {B}(H_2,K_1)$ such that
$$\psi(A\otimes B)=\frac{M_1BM_1^*}{{\rm
Tr}(M_1BM_1^*)}\otimes R_2 \ \mbox{\rm for all } A \in {\mathcal S}
(H_1)  \ \mbox{\rm and}\  B \in {\mathcal S} (H_2)$$ or $$
\psi(A\otimes B)=\frac{M_1B^tM_1^*}{{\rm Tr}(M_1B^tM_1^*)}\otimes
R_2 \ \mbox{\rm for all } A \in {\mathcal S} (H_1)  \ \mbox{\rm
and}\  B \in {\mathcal S} (H_2).$$}

(5) {\it $\dim H_1\leq \dim K_2$, there exist $R_1\in \mathcal
{P}ur(K_1)$ and injective $M_2\in \mathcal {B}(H_1, K_2)$ such that
$$\psi(A\otimes B)=R_1\otimes \frac{M_2AM_2^*}{{\rm
Tr}(M_2AM_2^*)} \ \mbox{\rm for all } A \in {\mathcal S} (H_1)  \
\mbox{\rm and}\  B \in {\mathcal S} (H_2)$$ or $$ \psi(A\otimes
B)=R_1\otimes \frac{M_2A^tM_2^*}{{\rm Tr}(M_2A^tM_2^*)} \ \mbox{\rm
for all } A \in {\mathcal S} (H_1)  \ \mbox{\rm and}\  B \in
{\mathcal S} (H_2).$$}

(6) {\it $\dim H_i\leq \dim K_i$, $i=1,2$, there exist injective
$M_1\in \mathcal {B}(H_1, K_1)$ and $M_2\in \mathcal {B}(H_2, K_2)$
such that
$$\psi(A\otimes B)=\frac{M_1\Psi_1(A)M_1^*}{{\rm Tr}(M_1\Psi_1(A)M_1^*)}\otimes \frac{M_2\Psi_2(B)M_2^*}{{\rm Tr}(M_2\Psi_2(B)M_2^*)}$$
for all $A \in {\mathcal S} (H_1)  \ \mbox{\rm and}\  B \in {\mathcal S} (H_2)$, where
$\Psi_i:{\mathcal B}(H_i)\rightarrow {\mathcal B}(H_i)$, $i=1, 2$,
is the identity, or the transpose.}

(7) {\it $\dim H_1\leq\dim K_2$ and $\dim H_2\leq\dim K_1$, there
exist injective $M_1\in \mathcal {B}(H_2, K_1)$ and $M_2\in \mathcal
{B}(H_1, K_2)$ such that
$$\psi(A\otimes B)=\frac{M_1\Psi_2(B)M_1^*}{{\rm Tr}(M_1\Psi_2(B)M_1^*)}\otimes \frac{M_2\Psi_1(A)M_2^*}{{\rm Tr}(M_2\Psi_1(A)M_2^*)}$$
for all $A \in {\mathcal S} (H_1)  \ \mbox{\rm and}\  B \in {\mathcal S} (H_2)$, where
$\Psi_i:{\mathcal B}(H_i)\rightarrow {\mathcal B}(H_i)$, $i=1, 2$,
is the identity, or the transpose.}

(8) {\it $\min\{\dim H_1,\dim H_2\}\leq \dim K_1$, there exist
$R_2\in {\mathcal Pur}(K_2)$ and a strict convex combination
preserving map $\varphi_1: {\mathcal S}_{\rm sep}(H_1\otimes
H_2)\rightarrow {\mathcal S}(K_1)$ such that
$$\psi(A\otimes B)=\varphi_1(A\otimes B)\otimes R_2$$
 for all $A \in \mathcal {S} (H_1 )$ and $B \in \mathcal {S} (H_2 )$.
Moreover, $\varphi_1$ satisfies that
 for each $P\otimes Q\in \mathcal {P}ur(H_1)\otimes
\mathcal {P}ur(H_2)$, $\varphi_1(P\otimes Q)=\frac {M_PQM_P^*}{{\rm
Tr}(M_PQM_P^*)}=\frac {N_QPN_Q^*}{{\rm Tr}(N_QPN_Q^*)}$  for some
injective linear or conjugate linear  (may not synchronously)
 operators $M_P: H_2\rightarrow K_1$ and $N_Q: H_1\rightarrow K_1$.}

(9) {\it $\min\{\dim H_1,\dim H_2\}\leq \dim K_2$, there exist
$R_1\in \mathcal {P}ur(K_1)$ and a strict convex combination
preserving map $\varphi_2: {\mathcal S}_{\rm sep}(H_1\otimes
H_2)\rightarrow {\mathcal S}(K_2)$ such that
$$\psi(A\otimes B)=R_1\otimes \varphi_2(A\otimes B)$$  for all
$A \in \mathcal {S} (H_1 )$ and $B \in \mathcal {S} (H_2 )$.
Moreover, $\varphi_2$ satisfies that, for each $P\otimes Q\in
\mathcal {P}ur(H_1)\otimes \mathcal {P}ur(H_2)$, $\varphi_2(P\otimes
Q)=\frac {M_PQM_P^*}{{\rm Tr}(M_PQM_P^*)}=\frac {N_QPN_Q^*}{{\rm
Tr}(N_QPN_Q^*)}$, for some injective linear or conjugate linear (may
not synchronously) operators $M_P: H_2\rightarrow K_2$ and $N_Q:
H_1\rightarrow K_2$.}

(10) {\it There exist $P_i'\in{\mathcal Pur}(K_1)$ and
$Q_i'\in{\mathcal Pur}(K_2)$, $i=1,2$ such that $\psi({\mathcal
Pur}(H_1)\otimes{\mathcal Pur}(H_2))=\{P_1'\otimes Q_1',P_2'\otimes
Q_2'\}$, and {\rm ran}$(\psi)\subseteq [P_1'\otimes Q_1',P_2'\otimes
Q_2']$. \\
Where the transpose is taken with respect to an arbitrarily fixed
orthonormal basis.}

{\bf Proof.}  Suppose $\dim H_1=m$, $\dim H_2=n$. If the range of
$\psi$ is collinear, then it is clear that either (1) holds or (10)
holds. So, in the sequel, we assume that the range of $\psi$ is
non-collinear. Denote by $\mathcal {B}_{\rm sa}(H)$ the real linear
space of all self-adjoint operators on the Hilbert space $H$.
Consider the partial traces ${\rm Tr_1}:\mathcal{B}_{\rm
sa}(K_1\otimes K_2)\rightarrow \mathcal{B}_{\rm sa}(K_2)$ and ${\rm
Tr_2}:\mathcal{B}_{\rm sa}(K_1\otimes K_2)\rightarrow
\mathcal{B}_{\rm sa}(K_1)$ on $\mathcal{B}_{\rm sa}(K_1\otimes
K_2)\equiv \mathcal{B}_{\rm sa}(K_1)\otimes \mathcal{B}_{\rm
sa}(K_2)$ defined by ${\rm Tr_1}(A\otimes B)=({\rm Tr}A)B$ and ${\rm
Tr_2}(A\otimes B)=({\rm Tr}B)A$. Clearly, ${\rm Tr_1}$ and ${\rm
Tr_2}$ are linear maps. Define two maps $\phi_1:(\mathcal
{S}(H_1),\mathcal {S}(H_2))\rightarrow \mathcal {S}(K_1)$ and
$\phi_2:(\mathcal {S}(H_1),\mathcal {S}(H_2))\rightarrow \mathcal
{S}(K_2)$ by
$$\phi_1(A,B)={\rm Tr_2}(\psi(A\otimes B)) \ {\rm and} \ \phi_2(A,B)={\rm Tr_1}(\psi(A\otimes
B)).$$ Notice that $$\psi(P\otimes Q)=\phi_1(P,Q)\otimes \phi_2(P,Q)
\ {\rm for \ all} \ P\in \mathcal {P}ur(H_1) \ {\rm and} \ Q\in
\mathcal {P}ur(H_2).$$ Fix a $Q\in \mathcal {P}ur(H_2)$; then the
maps $\phi_1(\cdot,Q):\mathcal {S}(H_1)\rightarrow \mathcal
{S}(K_1)$ and $\phi_2(\cdot,Q):\mathcal {S}(H_1)\rightarrow \mathcal
{S}(K_2)$ are both strict convex combination  preserving and
$\phi_1(\mathcal {P}ur(H_1),Q)\subseteq \mathcal {P}ur(K_1)$ while
$\phi_2(\mathcal {P}ur(H_1),Q)\subseteq \mathcal {P}ur(K_2)$.
Therefore, applying Theorem 2.5 to $\phi_1(\cdot,Q)$ and
$\phi_2(\cdot,Q)$, respectively, we get that, for $i=1,2$, either

(i) there exists pure state $R_{iQ}\in{\mathcal Pur}(K_i)$ such that
$\phi_i(A, Q)=R_{iQ}$ for all $A\in \mathcal {S}(H_1)$; \\ or

(ii) there are pure states $R_{1Q}^{(i)}, R_{2Q}^{(i)}\in{\mathcal
Pur}(K_i)$ and a strict convex combination preserving map
$h_{iQ}:{\mathcal S}(H_1)\rightarrow [0, 1]$ such that
$\phi_i({\mathcal Pur}(H_1),Q)=\{R_{1Q}^{(i)}, R_{2Q}^{(i)}\}$ and
$\phi_i(A, Q)=h_{iQ}(A)R_{1Q}^{(i)}+(1-h_{iQ}(A))R_{2Q}^{(i)}$ for
all $A\in \mathcal {S}(H_1)$;\\ or

(iii) there exists an injective   linear or conjugate linear
operator $M_{iQ}:H_1\rightarrow K_i$ such that $\phi_i(A,
Q)=\frac{M_{iQ}AM_{iQ}^*}{{\rm Tr}(M_{iQ}AM_{iQ}^*)}$ for all $A\in
\mathcal {S}(H_1)$.

 As $\psi$ is
strict convex combination preserving and ${\rm ran}\psi$ is
non-collinear, Lemma 2.3 is applicable. Thus we have
$\psi(\rho)=\frac{\Gamma(\rho)+D}{f(\rho)+d}$, where $\Gamma:
\mathcal {B}_{\rm sa}(H_1\otimes H_2)\rightarrow \mathcal {B}_{\rm
sa}(K_1\otimes K_2)$ is a linear map, $D\in \mathcal {B}_{\rm
sa}(K_1\otimes K_2)$ is an operator, $f:\mathcal {B}_{\rm
sa}(H_1\otimes H_2)\rightarrow \mathbb{R}$ is a linear functional
and $d\in \mathbb{R}$ with $f(\rho)+d>0$ for all $\rho\in \mathcal
{S}_{\rm sep}(H_1\otimes H_2)$. As both $H_1$ and $H_2$ are finite
dimensional, $\Gamma$, $f$ are continuous. It follows that $\psi$ is
continuous. So both $\phi_1$ and $\phi_2$ are continuous on
$\mathcal {S}(H_1)$. These facts will be used frequently.

Let us first consider the map $\phi_1$.

{\bf Claim 1.} Either $\phi_1(\cdot,Q)$ has the form (i) for all
$Q\in \mathcal {P}ur(H_2)$ or $\phi_1(\cdot,Q)$ has the form (iii)
for all $Q\in \mathcal {P}ur(H_2)$.

As mentioned above, for any fixed pure state $Q\in{\mathcal
Pur}(H_2)$, $\phi_1(\cdot, Q)$ takes one of the forms (i)-(iii). \if
We assert that $\phi_1(\cdot,Q)$ can not have the form (ii) for any
$Q\in \mathcal {P}ur(H_2)$. If, on the contrary,  there exists some
$Q_0\in \mathcal {P}ur(H_2)$ such that $\phi_1(\cdot,Q_0)$ has the
form (ii). Then $\phi_1({\mathcal Pur}(H_1), Q_0)=\{R_{1Q_0}^{(1)},
R_{2Q_0}^{(1)}\}$. Let $P_1=x_1\otimes x_1$, $P_2=x_2\otimes x_2\in
\mathcal {P}ur(H_1)$ such that $\phi_1(P_1,Q_0)=R_{1Q_0}^{(1)}$ and
$\phi_1(P_2,Q_0)=R_{2Q_0}^{(1)}$. Let $P(\alpha, \beta)=(\alpha
x_1+\beta x_2)\otimes (\alpha x_1+\beta x_2)$, where
$|\alpha|^2+|\beta|^2=1$. Then $P(\alpha, \beta)\in \mathcal
{P}ur(H_1)$ and  $f(\alpha,\beta)=\phi_1(P(\alpha, \beta),Q_0)\in
\{R_{1Q_0}^{(1)}, R_{2Q_0}^{(1)}\}$. Since $f(\alpha,\beta)$ is
continuous in $(\alpha,\beta)$, we see that $f(\alpha,\beta)$ can
only take one value, which is a contradiction.\fi As $\phi_1$ is
continuous, by Remark 2.7 we see that   $\phi_1(\cdot,Q)$ can not
have the form (ii) for any $Q\in \mathcal {P}ur(H_2)$. Thus, for any
$Q$, $\phi_1(\cdot,Q)$ takes the form (i) or the form (iii).

Furthermore,  we will show that either $\phi_1(\cdot,Q)$ has the
form (i) for all $Q\in \mathcal {P}ur(H_2)$ or $\phi_1(\cdot,Q)$ has
the form (iii) for all $Q\in \mathcal {P}ur(H_2)$.

To do this, for any $A\in {\mathcal S}(H_1)$ so that ${\rm
rank}(A)\geq2$, define $F_A:\mathcal {P}ur(H_2)\rightarrow \Bbb{R}$
by $F_A(Q)=\|\phi_1(A,Q)\|_2$, where $\|\cdot\|_2$ is the
Hilbert-Schmidt norm. Notice that when $\phi_1(\cdot,Q)$ takes the
form (i), then $F_A(Q)=\|\phi_1(A,Q)\|_2=1$; when $\phi_1(\cdot,Q)$
takes the form (iii), then $\phi_1(A,Q)=\frac {M_{1Q}AM_{1Q}^*}{{\rm
Tr}(M_{1Q}AM_{1Q}^*)}$ and $F_A(Q)=\|\phi_1(A,Q)\|_2=\frac{\|
M_{1Q}AM_{1Q}^*\|_2}{\| M_{1Q}A^{\frac 12}\|_2^2}<1$ as ${\rm rank}
A\geq 2$ and $M_{1Q}$ is injective.

If there exist two distinct $Q_1, Q_2\in \mathcal {P}ur(H_2)$, such
that $\phi_1(\cdot, Q_1)$ has the form (i) while $\phi_1(\cdot,
Q_2)$ has the form (iii), that is, $\phi_1(\cdot,Q_1)=R_{1Q_1}$,
$\phi_1(\cdot,Q_2)=\frac{M_{1Q_2}(\cdot)M_{1Q_2}^*}{{\rm
Tr}(M_{1Q_2}(\cdot)M_{1Q_2}^*)}$. Let $Q_1=x\otimes x$ and
$Q_2=y\otimes y$ with unit vectors $x, y\in H_2\cong\mathbb{C}^n$.
Note that $x$ and $y$ are linearly independent. For any $t\in
[0,1]$, define
$$Q(t)=\frac{1}{\|x+t(y-x)\|^2} (x+t(y-x))\otimes (x+t(y-x))\in \mathcal
{P}ur(H_2).$$ Then, $Q(t)$ is continuous in $t$, $Q(0)=Q_1$,
$Q(1)=Q_2$, and $\phi_1(\cdot,Q(t))$
  has the form (i) or (iii) for any $t$. Fix an $A\in{\mathcal S}(H_1)$ with rank$A\geq 2$ and let $t_0=\max\{t\in
[0,1]:F_A(Q(t))=1\}$. Then $F_A(Q(t_0))=1$ and so $\phi_1(\cdot,
Q(t_0))$ has the form (i). Thus $\|\phi_1(A, Q(t_0))\|_2=1$ for any
$A\in {\mathcal S}(H_1)$. For any $1\geq t>t_0$, $\phi_1(\cdot,
Q(t))$ has the form (iii). Thus there exist $\{t_n\}$, $t_n>t_0$,
such that $\phi_1(\cdot, Q(t_n))$ has the form (iii) and
$t_n\rightarrow t_0$. Then by Lemma 3.1, for any given sufficient
small $\varepsilon>0$, there exist $\{A_{t_n}\}\subseteq {\mathcal
S}(H_1)$ with ${\rm rank}A_{t_n}=2$, such that $\frac 12\leq
\|\phi_1(A_{t_n},Q(t_n))\|_2^2\leq \frac 12+\varepsilon<1$. The
reason of the existence of $\{A_{t_n}\}$  for each $n$ is that,  we
can find rank-2 operator $B_{t_n}\in$ran$\phi_1(\cdot, Q_{t_n})$
such that $\frac12 \leq\|B_{t_n}\|_2^2\leq \frac 12+\varepsilon<1$,
thus there exists $A_{t_n}\in{\mathcal S}(H_1)$ such that
$B_{t_n}=\frac{M_{t_n}A_{t_n}M_{t_n}^*}{{\rm
Tr}(M_{t_n}A_{t_n}M_{t_n}^*)}$. As $M_{t_n}$ is injective, we see
that rank$A_{t_n}=2$. Now, since ${\mathcal S}(H_1)$ is a compact
set, $\{A_{t_n}\}\subseteq {\mathcal S}(H_1)$ has a convergent
subsequence
  $\{A_{t_{n_i}}\}\subseteq \{A_{t_n}\}$, say
$A_{t_{n_i}}\rightarrow A_0$ as $t_{n_i}\rightarrow t_0$. Then the
continuity of $\psi$ entails that $\|\phi_1(A_{t_{n_i}},
Q(t_{n_i}))\|_2^2\rightarrow \|\phi_1(A_0,Q(t_0))\|_2^2$. But this
is a contradiction because $\|\phi_1(A_0,Q(t_0))\|_2^2=1$ and
$\|\phi_1(A_{t_{n_i}}, Q(t_{n_i}))\|_2^2\leq \frac
12+\varepsilon<1$. Thus either $\phi_1(\cdot,Q)$ has the form (i)
for all $Q\in \mathcal {P}ur(H_2)$ or $\phi_1(\cdot,Q)$ has the form
(iii) for all $Q\in \mathcal {P}ur(H_2)$.

Similarly, we have

{\bf Claim 1$^\prime$.} Either $\phi_2(\cdot,Q)$ has the form (i)
for all $Q\in \mathcal {P}ur(H_2)$ or $\phi_2(\cdot,Q)$ has the form
(iii) for all $Q\in \mathcal {P}ur(H_2)$.

{\bf Claim 2.} One of the following holds:

(a) For all $Q\in \mathcal {P}ur(H_2)$, both $\phi_1(\cdot,Q)$ and
$\phi_2(\cdot,Q)$ have the form (i).

(b) For all $Q\in \mathcal {P}ur(H_2)$, $\phi_1(\cdot,Q)$ has the
form (i) and $\phi_2(\cdot,Q)$ has the form (iii).

(c) For all $Q\in \mathcal {P}ur(H_2)$, $\phi_1(\cdot,Q)$ has the
form (iii) and $\phi_2(\cdot,Q)$ has the form (i).

We need only to check that, for all $Q\in \mathcal {P}ur(H_2)$,
$\phi_1(\cdot, Q)$ and $\phi_2(\cdot, Q)$ can not have the form
(iii) simultaneously. Suppose, on the contrary, there exists $Q_0\in
\mathcal {P}ur(H_2)$ such that both $\phi_1(\cdot,Q_0)$ and
$\phi_2(\cdot,Q_0)$ are of the form (iii). Then there exist
injective linear or conjugate linear  operators
$M_{1Q_0}:H_1\rightarrow K_1$ and $M_{2Q_0}:H_1\rightarrow K_2$ such
that $\phi_i(A,Q_0)= \frac{M_{iQ_0}AM_{iQ_0}^*}{{\rm
Tr}(M_{iQ_0}AM_{iQ_0}^*)}$ for all $A\in {\mathcal S}(H_1)$, $i=1,
2$. Thus, we must have $\dim H_1\leq \min\{\dim K_1,\dim K_2\}$ and
$$\begin{array}{rl} \psi(P\otimes Q_0)=&\phi_1(P,Q_0)\otimes
\phi_2(P,Q_0)=(\frac {M_{1Q_0}PM_{1Q_0}^*}{{\rm
Tr}(M_{1Q_0}PM_{1Q_0}^*)})\otimes (\frac{M_{2Q_0}PM_{2Q_0}^*}{{\rm
Tr}(M_{2Q_0}PM_{2Q_0}^*)})\\
=&\frac{(M_{1Q_0}\otimes M_{2Q_0})(P\otimes P)(M_{1Q_0}\otimes
M_{2Q_0})^*}{{\rm Tr}((M_{1Q_0}\otimes M_{2Q_0})(P\otimes
P)(M_{1Q_0}\otimes M_{2Q_0})^*)}
\end{array}$$ for all $P\in \mathcal {P}ur(H_1)$. Particularly, take $P_1=e_1\otimes e_1$,
$P_2=e_2\otimes e_2$, $P_3=\frac12 (e_1\otimes e_1+e_1\otimes
e_2+e_2\otimes e_1+e_2\otimes e_2)$ and $P_4=\frac12 (e_1\otimes
e_1-e_1\otimes e_2-e_2\otimes e_1+e_2\otimes e_2)$, where
$e_1,e_2\in H_1$ are unit vectors with $e_1\perp e_2$. Then
$P_1+P_2=P_3+P_4$ and so $P_1\otimes Q_0+P_2\otimes Q_0=P_3\otimes
Q_0+P_4\otimes Q_0$. As $\psi$ is strict convex combination
preserving, there exist $s_1\in(0,1)$ and $s_2\in(0,1)$, such that
$$\begin{array}{rl} &\psi(\frac12 P_1\otimes Q_0+\frac12 P_2\otimes
Q_0)=s_1\psi(P_1\otimes Q_0)+(1-s_1)\psi(P_2\otimes Q_0)\\
=&s_1\frac{(M_{1Q_0}\otimes M_{2Q_0})(P_1\otimes
P_1)(M_{1Q_0}\otimes M_{2Q_0})^*}{{\rm Tr}((M_{1Q_0}\otimes
M_{2Q_0})(P_1\otimes P_1)(M_{1Q_0}\otimes
M_{2Q_0})^*)}+(1-s_1)\frac{(M_{1Q_0}\otimes M_{2Q_0})(P_2\otimes
P_2)(M_{1Q_0}\otimes M_{2Q_0})^*}{{\rm Tr}((M_{1Q_0}\otimes
M_{2Q_0})(P_2\otimes P_2)(M_{1Q_0}\otimes M_{2Q_0})^*)}
\end{array}$$ and $$\begin{array}{rl} &\psi(\frac12 P_3\otimes Q_0+\frac12
P_4\otimes
Q_0)=s_2\psi(P_3\otimes Q_0)+(1-s_2)\psi(P_4\otimes Q_0)\\
=&s_2\frac{(M_{1Q_0}\otimes M_{2Q_0})(P_3\otimes
P_3)(M_{1Q_0}\otimes M_{2Q_0})^*}{{\rm Tr}((M_{1Q_0}\otimes
M_{2Q_0})(P_3\otimes P_3)(M_{1Q_0}\otimes
M_{2Q_0})^*)}+(1-s_2)\frac{(M_{1Q_0}\otimes M_{2Q_0})(P_4\otimes
P_4)(M_{1Q_0}\otimes M_{2Q_0})^*}{{\rm Tr}((M_{1Q_0}\otimes
M_{2Q_0})(P_4\otimes P_4)(M_{1Q_0}\otimes M_{2Q_0})^*)}.
\end{array}$$
It follows that there exist $a_i\neq 0$, $i=1, 2, 3, 4$, such that
$$a_1P_1\otimes P_1+a_2P_2\otimes P_2=a_3P_3\otimes P_3+a_4P_4\otimes
P_4.$$ \if In fact, $a_1=\frac{s_1}{{\rm Tr}((M_{1Q_0}\otimes
M_{2Q_0})(P_1\otimes P_1)(M_{1Q_0}\otimes M_{2Q_0})^*)}$,
$a_2=\frac{1-s_1}{{\rm Tr}((M_{1Q_0}\otimes M_{2Q_0})(P_2\otimes
P_2)(M_{1Q_0}\otimes M_{2Q_0})^*)}$, $a_3=\frac{s_2}{{\rm
Tr}((M_{1Q_0}\otimes M_{2Q_0})(P_3\otimes P_3)(M_{1Q_0}\otimes
M_{2Q_0})^*)}$, $a_4=\frac{1-s_2}{{\rm Tr}((M_{1Q_0}\otimes
M_{2Q_0})(P_4\otimes P_4)(M_{1Q_0}\otimes M_{2Q_0})^*)}$.\fi  As
$P_4=P_1+P_2-P_3$, then $0=a_1P_1\otimes P_1+a_2P_2\otimes
P_2-a_3P_3\otimes P_3-a_4P_4\otimes P_4=P_1\otimes a_1P_1+P_2\otimes
a_2P_2-P_3\otimes a_3P_3-P_1\otimes a_4P_4-P_2\otimes
a_4P_4+P_3\otimes a_4P_4=P_1\otimes (a_1P_1-a_4P_4)+P_2\otimes
(a_2P_2-a_4P_4)-P_3\otimes (a_3P_3-a_4P_4)$. As $P_1$, $P_2$ and
$P_3$ are linearly independent, by \cite{H3}, we have
$a_1P_1-a_4P_4=0$, $a_2P_2-a_4P_4=0$ and $a_3P_3-a_4P_4=0$, which is
a contradiction. So the claim is true.

Similarly, one can check that

{\bf Claim 3.} One of the following holds:

(a$^\prime$) For all $P\in \mathcal {P}ur(H_1)$, both
$\phi_1(P,\cdot)$ and $\phi_2(P,\cdot)$ have the form (i).

(b$^\prime$) For all $P\in \mathcal {P}ur(H_1)$, $\phi_1(P,\cdot)$
has the form (i) and $\phi_2(P,\cdot)$ has the form (iii).

(c$^\prime$) For all $P\in \mathcal {P}ur(H_1)$, $\phi_1(P,\cdot)$
has the form (iii) and $\phi_2(P,\cdot)$ has the form (i).

{\bf Claim 4.}  (a) and (a$^\prime$) can not hold simultaneously.

In fact, if (a) and (a$^\prime$) hold,  that is, for all $Q\in
\mathcal {P}ur(H_2)$, we have $\phi_i(A, Q)=R_{iQ}$, and, for all
$P\in \mathcal {P}ur(H_1)$, we have $\phi_i(P, B)=R_{iP}$. Fix
$P_0\in \mathcal {P}ur(H_1)$ and $Q_0\in \mathcal {P}ur(H_2)$. Then
we get
$$\phi_i(P, Q)=\phi_i(P,Q_0)=\phi_i(P_0,Q_0)=R_i.$$ Therefore, $\psi(P\otimes Q)=\phi_1(P,Q)\otimes \phi_2(P,Q)=R_1\otimes
R_2$ for all $P\otimes Q\in \mathcal {P}ur(H_1)\otimes \mathcal
{P}ur(H_2)$. We know that for any $A\otimes B\in {\mathcal
S}(H_1)\otimes {\mathcal S}(H_2)$, there exist
$\{A_i\}_{i=1}^m\subseteq\mathcal {P}ur(H_1)$, $\{B_j\}_{j=1}^n
\subseteq \mathcal {P}ur(H_2)$, $\{a_i\}_{i=1}^m\subseteq[0,1]$,
$\{b_j\}_{j=1}^n \subseteq [0,1]$ satisfying $\sum_{i=1}^m a_i=1$,
$\sum_{j=1}^n b_j=1$, such that $A=\sum_{i=1}^m a_iA_i$,
$B=\sum_{j=1}^n b_jB_j$. Then, there exist $c_{ij}\geq 0$ with
$\sum_{i,j}c_{ij}=1$ such that $\psi(A\otimes
B)=\sum_{i,j}c_{ij}\psi(A_i\otimes B_j)=R_1\otimes R_2$, this
contradicts to the assumption that the range of $\psi$ is
non-collinear. So the Claim 4 is true.

{\bf Claim 5.} If (a) and (b$^\prime$) hold, then $\psi$ has the
form (3), that is, there exist $R_1\in \mathcal {P}ur(K_1)$ and
injective linear or conjugate linear operator $M_2:H_2\rightarrow
K_2 $ such that
$$\psi(A\otimes B)=R_1\otimes \frac{M_2BM_2^*}{{\rm
Tr}(M_2BM_2^*)} $$  for all  $ A \in {\mathcal S} (H_1)$   and $B
\in {\mathcal S} (H_2)$.   It is clear that $\dim H_2\leq \dim K_2$.

In this case, for any $P\otimes Q\in \mathcal {P}ur(H_1)\otimes
\mathcal {P}ur(H_2)$ we have $$\psi(P\otimes Q)=\phi_1(P,Q)\otimes
\phi_2(P,Q)=R_{1Q}\otimes R_{2Q}=R_{1P}\otimes
\frac{M_{2P}QM_{2P}^*}{{\rm Tr}(M_{2P}QM_{2P}^*)},$$ which implies
that $R_{1Q}=R_{1P}$ is independent of $P,Q$, and
$\frac{M_{2P}QM_{2P}^*}{{\rm Tr}(M_{2P}QM_{2P}^*)}=R_{2Q}$. Thus for
any fixed $Q\in \mathcal {P}ur(H_2)$,
$\frac{M_{2P_1}QM_{2P_1}^*}{{\rm
Tr}(M_{2P_1}QM_{2P_1}^*)}=\frac{M_{2P_2}QM_{2P_2}^*}{{\rm
Tr}(M_{2P_2}QM_{2P_2}^*)}=R_{2Q}$ holds for any distinct $P_1\in
\mathcal {P}ur(H_1)$ and $P_2\in \mathcal {P}ur(H_1)$. Thus for all
$Q\in \mathcal {P}ur(H_2)$, we have $\frac{M_{2P_1}QM_{2P_1}^*}{{\rm
Tr}(M_{2P_1}QM_{2P_1}^*)}=\frac{M_{2P_2}QM_{2P_2}^*}{{\rm
Tr}(M_{2P_2}QM_{2P_2}^*)}$, that is, $\frac{M_{2P}QM_{2P}^*}{{\rm
Tr}(M_{2P}QM_{2P}^*)}$ is independent of $P$. So there exist $R_1\in
\mathcal {P}ur(K_1)$ and injective   linear or conjugate linear
operator $M_2:H_2\rightarrow K_2$ such that $\psi(P\otimes
Q)=R_1\otimes \frac{M_2QM_2^*}{{\rm Tr}(M_2QM_2^*)}$ for all
separable pure states $P\otimes Q$.

Now, for any $B\in{\mathcal S}(H_2)$ and $P\in{\mathcal Pur}(H_1)$,
writing $B= \sum_{j=1}^n b_jB_j$ as in the proof of Claim 4, we have
$\psi(P\otimes B)=\psi(P\otimes \sum_{j=1}^n b_jB_j)=\sum_{j=1}^n
b'_j\psi(P\otimes B_j)=\sum_{j=1}^n b'_j(R_1\otimes
\frac{M_2B_jM_2^*}{{\rm Tr}(M_2B_jM_2^*)})=R_1\otimes \sum_{j=1}^n
b'_j\frac{M_2B_jM_2^*}{{\rm Tr}(M_2B_jM_2^*)}$ is a product state.
But we already know that
$\phi_2(P,\cdot)=\frac{M_2(\cdot)M_2^*}{{\rm Tr}(M_2(\cdot)M_2^*)}$,
so ${\rm Tr}_1(\psi(P\otimes B))=\phi_2(P,B)=\frac{M_2BM_2^*}{{\rm
Tr}(M_2BM_2^*)}$. Thus $\psi(P\otimes B)=R_1\otimes
\frac{M_2BM_2^*}{{\rm Tr}(M_2BM_2^*)}$. Then, for any $A\in{\mathcal
S}(H_1)$, writing $A=\sum _{i=1}^ma_iA_i$ as in the proof of Claim
4, we obtain $\psi(A\otimes B)=\psi(\sum_{i=1}^m a_iA_i\otimes
B)=\sum_{i=1}^m a'_i\psi(A_i\otimes B)=\sum_{i=1}^m a'_i (R_1\otimes
\frac{M_2BM_2^*}{{\rm Tr}(M_2BM_2^*)})=R_1\otimes
(\sum_{i=1}^ma'_i\frac{M_2BM_2^*}{{\rm Tr}(M_2BM_2^*)}))$, which is
a product states. Therefore, we must have $\psi(A\otimes
B)=R_1\otimes \frac{M_2BM_2^*}{{\rm Tr}(M_2BM_2^*)}$, where $M_2$ is
linear or conjugate linear. In the case that $M_2$ is a conjugate
linear operator,  it is well known that,  there exists a linear
operator $N_2$ such that $\frac{M_2BM_2^*}{{\rm
Tr}(M_2BM_2^*)}=\frac{N_2B^tN_2^*}{{\rm Tr}(N_2B^tN_2^*)}$ for all
$B$, where the transpose is taken with respect to   an arbitrarily
fixed orthonormal basis of $H_2$. So the claim is true.

Similarly, one can show the following Claims 6-8.

{\bf Claim 6.} If (a) and (c$^\prime$) hold, then (4) holds.

{\bf Claim 7.} If (b) and (a$^\prime$) hold, then $\psi$ has the
form (5).

{\bf Claim 8.} If (c) and (a$^\prime$) hold, then  $\psi$ takes the
form (2).

{\bf Claim 9.} If (b) and (c$^\prime$) hold, then  $\psi$ has the
form (7).

Suppose that (b) and (c$^\prime$) hold; then
$\phi_1(P,Q)=\phi_1(P_0,Q)=\frac{M_{1P_0}QM_{1P_0}^*}{{\rm
Tr}(M_{1P_0}QM_{1P_0}^*)}$ and
$\phi_2(P,Q)=\phi_2(P,Q_0)=\frac{M_{2Q_0}PM_{2Q_0}^*}{{\rm
Tr}(M_{2Q_0}PM_{2Q_0}^*)}$. Thus, we obtain $$\psi(P\otimes
Q)=\phi_1(P,Q)\otimes \phi_2(P,Q)=\frac{M_{1P_0}QM_{1P_0}^*}{{\rm
Tr}(M_{1P_0}QM_{1P_0}^*)}\otimes \frac{M_{2Q_0}PM_{2Q_0}^*}{{\rm
Tr}(M_{2Q_0}PM_{2Q_0}^*)}$$ for all $P\in \mathcal {P}ur(H_1)$ and
$Q\in \mathcal {P}ur(H_2)$, where $M_{1P_0}:H_2\rightarrow K_1$ and
$M_{2Q_0}:H_1\rightarrow K_2$ are injective linear or conjugate
linear operators. It follows that $\dim H_1\leq\dim K_2$ and $\dim
H_2\leq\dim K_1$. Let $M_1=M_{1P_0}$ and $M_2=M_{2Q_0}$. Then
$\psi(P\otimes Q)=\frac{M_1QM_1^*}{{\rm Tr}(M_1QM_1^*)}\otimes
\frac{M_2PM_2^*}{{\rm Tr}(M_2PM_2^*)}$ for all separable pure states
$P\otimes Q$. For any $A\otimes B\in {\mathcal S}(H_1\otimes H_2)$,
write $A=\sum_{i=1}^ma_iA_i$ and $B=\sum_{j=1}^nb_jB_j$ by the
spectral theorem, where $a_i,b_j\geq 0$ with $\sum_{i=1}^ma_i=1,
\sum_{j=1}^nb_j=1$ and $A_i$s, $B_j$s pure states. Thus, for any
$Q\in{\mathcal Pur}(H_2)$,  $\psi(A\otimes Q)=\psi(\sum_{i=1}^m
a_iA_i\otimes Q)=\sum_{i=1}^m a'_i\psi(A_i\otimes Q)=\sum_{i=1}^m
a'_i(\frac{M_1QM_1^*}{{\rm Tr}(M_1QM_1^*)}\otimes
\frac{M_2A_iM_2^*}{{\rm Tr}(M_2A_iM_2^*)})=\frac{M_1QM_1^*}{{\rm
Tr}(M_1QM_1^*)}\otimes (\sum_{i=1}^m a'_i\frac{M_2A_iM_2^*}{{\rm
Tr}(M_2A_iM_2^*)})$ is a product state. As $\phi_2(\cdot,
Q)=\frac{M_2(\cdot)M_2^*}{{\rm Tr}(M_2(\cdot)M_2^*)}$, we obtain
that ${\rm Tr}_1(\psi(A\otimes Q))=\phi_2(A,
Q)=\frac{M_2AM_2^*}{{\rm Tr}(M_2AM_2^*)}$. So we must have
$\psi(A\otimes Q)=\frac{M_1QM_1^*}{{\rm Tr}(M_1QM_1^*)}\otimes
\frac{M_2AM_2^*}{{\rm Tr}(M_2AM_2^*)}$. It follows that
$\psi(A\otimes B)=\psi(A\otimes \sum_{j=1}^n b_jB_j)=\sum_{j=1}^n
b'_j\psi(A\otimes B_j)=\sum_{j=1}^n b'_j (\frac{M_1B_jM_1^*}{{\rm
Tr}(M_1B_jM_1^*)}\otimes \frac{M_2AM_2^*}{{\rm
Tr}(M_2AM_2^*)})=(\sum_{j=1}^n b'_j \frac{M_1B_jM_1^*}{{\rm
Tr}(M_1B_jM_1^*)})\otimes \frac{M_2AM_2^*}{{\rm Tr}(M_2AM_2^*)}$. On
the other hand, we also have $\psi(P\otimes B)=\psi(P\otimes
\sum_{j=1}^nb_j B_j)=\sum_{j=1}^n b'_j\psi(P\otimes
B_j)=\sum_{j=1}^nb'_j (\frac{M_1B_jM_1^*}{{\rm
Tr}(M_1B_jM_1^*)}\otimes \frac{M_2PM_2^*}{{\rm
Tr}(M_2PM_2^*)})=(\sum_{j=1}^nb'_j \frac{M_1B_jM_1^*}{{\rm
Tr}(M_1B_jM_1^*)})\otimes \frac{M_2PM_2^*}{{\rm Tr}(M_2PM_2^*)}$. As
$\phi_1(P, \cdot)=\frac{M_1(\cdot)M_1^*}{{\rm
Tr}(M_1(\cdot)M_1^*)}$, then ${\rm Tr}_2(\psi(P\otimes B))=\phi_1(P,
B)=\frac{M_1BM_1^*}{{\rm Tr}(M_1BM_1^*)}$. Thus we get
$\psi(P\otimes B)=\frac{M_1BM_1^*}{{\rm Tr}(M_1BM_1^*)}\otimes
\frac{M_2PM_2^*}{{\rm Tr}(M_2PM_2^*)}$ and then   $\psi(A\otimes
B)=\psi(\sum_{i=1}^ma_iA_i\otimes B)=\sum_{i=1}^m a_i\psi(A_i\otimes
B)=\sum_{i=1}^m a'_i(\frac{M_1BM_1^*}{{\rm Tr}(M_1BM_1^*)}\otimes
\frac{M_2A_iM_2^*}{{\rm Tr}(M_2A_iM_2^*)})=\frac{M_1BM_1^*}{{\rm
Tr}(M_1BM_1^*)}\otimes(\sum_{i=1}^m a'_i\frac{M_2A_iM_2^*}{{\rm
Tr}(M_2A_iM_2^*)})$. Now it is clear that $\psi(A\otimes
B)=\frac{M_1BM_1^*}{{\rm Tr}(M_1BM_1^*)}\otimes
\frac{M_2AM_2^*}{{\rm Tr}(M_2AM_2^*)}$  for any $A\in{\mathcal
S}(H_1)$ and $B\in{\mathcal S}(H_2)$.   Hence the claim is true.

Similarly, we have

{\bf Claim 10.} If (c) and (b$^\prime$) hold, then  $\psi$ has the
form (6).

{\bf Claim 11.} If (b) and (b$^\prime$) hold, then $\psi$ has the
form (9).

Assume (b) and (b$^\prime$) hold synchronously. Then for any
$P\otimes Q\in \mathcal {P}ur(H_1)\otimes \mathcal {P}ur(H_2)$ we
have $\psi(P\otimes Q)=\phi_1(P,Q)\otimes \phi_2(P,Q)=R_{1Q}\otimes
\frac{M_{2Q}PM_{2Q}^*}{{\rm Tr}(M_{2Q}PM_{2Q}^*)}=R_{1P}\otimes
\frac{M_{2P}QM_{2P}^*}{{\rm Tr}(M_{2P}QM_{2P}^*)}$. It follows that
there exist $R_1\in \mathcal {P}ur(K_1)$ such that
$R_{1Q}=R_{1P}=R_1$ and $\frac{M_{2Q}PM_{2Q}^*}{{\rm
Tr}(M_{2Q}PM_{2Q}^*)}=\frac{M_{2P}QM_{2P}^*}{{\rm
Tr}(M_{2P}QM_{2P}^*)}$ for all $P$, $Q$. Thus there exists a strict
convex combination preserving  map $\varphi_2:{\mathcal S}_{\rm
sep}(H_1\otimes H_2)\rightarrow {\mathcal S}(K_2)$ such that, for
each $P\otimes Q\in {\mathcal Pur}(H_1)\otimes {\mathcal Pur}(H_2)$,
$\varphi_2(P\otimes Q)=\frac{M_PQM_P^*}{{\rm
Tr}(M_PQM_P^*)}=\frac{N_QPN_Q^*}{{\rm Tr}(N_QPN_Q^*)}$ for some
injective, may not synchronously, linear or conjugate linear
operators $M_P:H_2\rightarrow K_2$, $N_Q:H_1\rightarrow K_2$, and
$$\psi(\rho)=R_1\otimes \varphi_2(\rho)$$ for all $\rho\in {\mathcal S}_{\rm sep}(H_1\otimes
H_2)$. In this case $\max\{\dim H_1,\dim H_2\}\leq \dim K_2$. So
$\psi$ has the form (9) and Claim 11 is true.

Similarly,

{\bf Claim 12.} If (c) and (c$^\prime$) hold, then $\psi$ takes the
form (8).

\if Assume (c) and (c$^\prime$) hold synchronously. Then for any
$P\otimes Q\in \mathcal {P}ur(H_1)\otimes \mathcal {P}ur(H_2)$ we
have $\psi(P\otimes Q)=\phi_1(P,Q)\otimes
\phi_2(P,Q)=\frac{M_{1Q}PM_{1Q}^*}{{\rm Tr}(M_{1Q}PM_{1Q}^*)}\otimes
R_{2Q}=\frac{M_{1P}QM_{1P}^*}{{\rm Tr}(M_{1P}QM_{1P}^*)}\otimes
R_{2P}$. It follows that there exist $R_2\in \mathcal {P}ur(H_2)$
such that $R_{2Q}=R_{2P}=R_2$ and $\frac{M_{1Q}PM_{1Q}^*}{{\rm
Tr}(M_{1Q}PM_{1Q}^*)}=\frac{M_{1P}QM_{1P}^*}{{\rm
Tr}(M_{1P}QM_{1P}^*)}$ for all $P$, $Q$. Thus there exists a
strictly preserving convex combination map $\phi_1:S_{\rm
sep}(H_1\otimes H_2)\rightarrow S(H_1)$ such that, for each
$P\otimes Q\in \mathcal {P}ur(H_1)\otimes \mathcal {P}ur(H_2)$,
$\phi_1(P\otimes Q)=\frac{M_PQM_P^*}{{\rm
Tr}(M_PQM_P^*)}=\frac{N_QPN_Q^*}{{\rm Tr}(N_QPN_Q^*)}$ for some
injective and bounded, may not synchronously, linear or conjugate
linear operator $M_P:H_2\rightarrow H_1$, $N_Q:H_1\rightarrow H_1$,
and
$$\psi(\rho)=\phi_1(\rho)\otimes R_2$$ for all $\rho\in S_{\rm sep}(H_1\otimes
H_2)$. In this case $\dim H_1\geq \dim H_2$. So the claim is
true.\fi

Combining the claims 4-12, we complete the proof of Theorem 3.2.
\hfill$\Box$

By Theorem 3.2, the following corollary is immediate, which gives an
affirmative answer to a conjecture in \cite{HL}  without the
injectivity assumption.

\textbf{Corollary 3.3}  {\it Let $\psi:{\mathcal S}_{\rm
sep}(H_1\otimes H_2)\rightarrow {\mathcal S}_{\rm sep}(K_1\otimes
K_2)$ be a map with $2\leq\dim H_i<\infty$, $i=1, 2$, and ${\rm
ran}\psi$ non-collinear or a singleton (i.e., contains only one
element). If $\psi$ preserves separable pure states and strict
convex combinations, then it sends product states to product
states.}

Now we give a characterization of injective local quantum
measurements, which reveals that, in  almost all  situations the
maps preserving separable pure states and strict convex combinations
are essentially the injective local quantum measurements.

\textbf{Corollary 3.4} {\it Let $H_1, H_2, K_1,K_2$ be  Hilbert
spaces with $2\leq \dim H_i<\infty$, $i=1,2$ and let $\psi:{\mathcal
S}_{\rm sep}(H_1\otimes H_2)\rightarrow {\mathcal S}_{\rm
sep}(K_1\otimes K_2)$ be a map. Then the following statements are
equivalent.}

(1) {\it $\psi$ is strict convex combination preserving,
$\psi({\mathcal Pur}(H_1)\otimes {\mathcal Pur}(H_2))\subseteq
{\mathcal Pur}(K_1)\otimes {\mathcal Pur}(K_2)$ and the range of
$\psi$ is non-collinear   containing a state $\sigma $ so that both
reductions ${\rm Tr}_1(\sigma)$ and ${\rm Tr}_2(\sigma)$ have rank
$\geq 2$.}

(2) {\it $\psi$ is open line segment preserving, $\psi({\mathcal
Pur}(H_1)\otimes {\mathcal Pur}(H_2))\subseteq {\mathcal
Pur}(K_1)\otimes {\mathcal Pur}(K_2)$ and the range of $\psi$ is
non-collinear   containing a state $\sigma $ so that both reductions
${\rm Tr}_1(\sigma)$ and ${\rm Tr}_2(\sigma)$ have rank $\geq 2$.}

(3) {\it Either }

\hspace {4mm}(1$^\circ$) {\it there exist injective operators
$M_1\in \mathcal {B}(H_1, K_1)$ and $M_2\in \mathcal {B}(H_2,K_2)$
such that
$$\psi(\rho)=\frac{(M_1\otimes M_2)\Phi(\rho)(M_1\otimes M_2)^*}{{\rm Tr}((M_1\otimes M_2)\Phi(\rho)(M_1\otimes
M_2)^*)}$$ for all $\rho\in {\mathcal S}_{\rm sep}(H_1\otimes
H_2)$;\\ or}

 \hspace{4mm} (2$^\circ$) {\it  there exist injective operators $M_1\in \mathcal {B}(H_2,
K_1)$ and $M_2\in \mathcal {B}(H_1, K_2)$ such that
$$\psi(\rho)=\frac{(M_1\otimes M_2)\Phi(\Theta(\rho))(M_1\otimes M_2)^*}{{\rm Tr}((M_1\otimes M_2)\Phi(\Theta(\rho))(M_1\otimes
M_2)^*)}$$ for all $\rho\in {\mathcal S}_{\rm sep}(H_1\otimes H_2)$.\\
Here $\Phi$ is the identity, or the transpose, or the partial
transpose of the first system or the partial transpose of the second
system with respect to an arbitrarily fixed product basis, $\Theta$
is the swap.}

{\bf Proof.} (1)$\Leftrightarrow$(2)$\Leftarrow$(3) is obvious, we
need to check the (1)$\Rightarrow$(3).

Assume (1). It is clear that
 $\psi$ has one of the forms Theorem 3.2.(1)-(10) as $\psi$ satisfies all the conditions of Theorem 3.2. Furthermore, the assumption
that there exists $\sigma$ in ${\rm ran}\psi$ so that rank${\rm
Tr}_i(\sigma)\geq 2$, $i=1,2$ forces that $\psi$ can only take the
form (6) or (7), that is, $\psi(A\otimes
B)=\frac{M_1\Psi_1(A)M_1^*}{{\rm Tr}(M_1\Psi_1(A)M_1^*)}\otimes
\frac{M_2\Psi_2(B)M_2^*}{{\rm Tr}(M_2\Psi_2(B)M_2^*)}$ for all
$A\otimes B\in {\mathcal S}_{\rm sep}(H_1\otimes H_2)$, or
$\psi(A\otimes B)=\frac{M_1\Psi_2(B)M_1^*}{{\rm
Tr}(M_1\Psi_2(B)M_1^*)}\otimes \frac{M_2\Psi_1(A)M_2^*}{{\rm
Tr}(M_2\Psi_1(A)M_2^*)}$ for all $A\otimes B\in {\mathcal S}_{\rm
sep}(H_1\otimes H_2)$, where $\Psi_i$ is the identity or the
transpose with respect to an arbitrarily fixed orthonormal basis,
$i=1,2$. Thus either

(i) $\psi(A\otimes B)=\frac{(M_1\otimes M_2)\Phi(A\otimes
B)(M_1\otimes M_2)^*}{{\rm Tr}(M_1\otimes M_2)\Phi(A\otimes
B)(M_1\otimes M_2)^*)}$ for all $A\otimes B\in {\mathcal S}_{\rm
sep}(H_1\otimes H_2)$; or

(ii) $\psi(A\otimes B)=\frac{(M_1\otimes M_2)\Phi(\Theta(A\otimes
B))(M_1\otimes M_2)^*}{{\rm Tr}(M_1\otimes M_2)\Phi(\Theta(A\otimes
B))(M_1\otimes M_2)^*)}$ for all $A\otimes B\in {\mathcal S}_{\rm
sep}(H_1\otimes H_2)$,\\
 where $\Phi$ is the identity, or the
transpose, or the partial transpose of the first system or the
partial transpose of the second system with respect to an
arbitrarily fixed product basis, $\Theta$ is the swap.

Let $$\Delta (\rho)=\frac{(M_1^{[-1]}\otimes
M_2^{[-1]})\psi(\Phi(\rho))(M_1^{[-1]}\otimes M_2^{[-1]})^*}{{\rm
Tr}((M_1^{[-1]}\otimes M_2^{[-1]})\psi(\Phi(\rho))(M_1^{[-1]}\otimes
M_2^{[-1]})^*)}
$$
if $\psi$ has form (i), and let
$$\Delta (\rho)=\frac{\Theta[(M_1^{[-1]}\otimes
M_2^{[-1]})\psi(\Phi(\rho))(M_1^{[-1]}\otimes M_2^{[-1]})^*]}{{\rm
Tr}((M_1^{[-1]}\otimes M_2^{[-1]})\psi(\Phi(\rho))(M_1^{[-1]}\otimes
M_2^{[-1]})^*)}
$$ if $\psi$ takes the form (ii). Then $\Delta :{\mathcal S}_{\rm
sep}(H_1\otimes H_2)\rightarrow {\mathcal S}_{\rm sep}(H_1\otimes
H_2)$ is a bijective map preserving separable pure states and strict
convex combinations. Furthermore $\Delta (A\otimes B)=A\otimes B$
for all $A\otimes B\in {\mathcal S}_{\rm sep}(H_1\otimes H_2)$. Then
by   \cite[Lemma 2.7]{HL}, we  get   $\Delta(\rho)=\rho$ for any
$\rho\in {\mathcal S}_{\rm sep}(H_1\otimes H_2)$. Now it is clear
that $\psi$ has the form (1$^\circ$) or (2$^\circ$), finishing the
proof. \hfill$\Box$

The following result is a  generalization of the main result in
\cite{HL} by omitting the additional assumption in Eq.(1.1)  for
finite dimensional case, and thus, answer affirmatively a conjecture
proposed in \cite{HL} , as mentioned in the introduction section.

\textbf{Theorem 3.5}   {\it Let $H_i, K_i,$ be complex Hilbert
spaces with $2\leq\dim H_i < \infty$, $i=1,2$, and
  $\psi: {\mathcal S}_{\rm sep} (H_1\otimes H_2)\rightarrow {\mathcal S}_{\rm sep}(K_1\otimes K_2)$ an bijective  map.
Then $\Phi$ is convex combination preserving if and only if either}

(1) {\it there exist invertible operators $S\in \mathcal
{B}(H_1,K_1)$ and $T\in \mathcal {B}(H_2,K_2)$ such that
$$\psi(\rho)=\frac{(S\otimes T)\Psi(\rho)(S\otimes T)^*}{{\rm Tr}((S\otimes T)\Psi(\rho)(S\otimes
T)^*)}$$ for all $\rho\in {\mathcal S}_{\rm sep}(H_1\otimes H_2)$;\\
or}

(2) {\it   there exist invertible operators $S\in \mathcal {B}(H_2,
K_1)$ and $T\in \mathcal {B}(H_1, K_2)$ such that
$$\psi(\rho)=\frac{(S\otimes T)\Psi(\Theta(\rho))(S\otimes T)^*}{{\rm Tr}((S\otimes T)\Psi(\Theta(\rho))(S\otimes
T)^*)}$$ for all $\rho\in {\mathcal S}_{\rm sep}(H_1\otimes H_2)$.\\
Here $\Psi$ is the identity, or the transpose, or the partial
transpose of the first system or the partial transpose of the second
system with respect to an arbitrarily fixed product basis, $\Theta$
is the swap. }

{\bf Proof.} We need only check the ``only if" part. By the
assumption, $\psi$ is bijective and strict convex combination
preserving from ${\mathcal S}_{\rm sep}(H_1\otimes H_2)$ onto
${\mathcal S}_{\rm sep}(K_1\otimes K_2) $. Particularly, the range
of $\psi$ is non-collinear.  By \cite{HL}, we have $\psi({\mathcal
Pur}_{\rm sep}(H_1\otimes H_2) )={\mathcal Pur}_{\rm sep}(K_1\otimes
K_2) $, that is, $\psi$ preserves separable pure states in both
directions.

Then by Corollary 3.4, $\psi$ has either the form (1) or the form
(2) with $T,S$ injective. Since $\psi({\mathcal Pur}_{\rm
sep}(H_1\otimes H_2) )={\mathcal Pur}_{\rm sep}(K_1\otimes K_2) $,
it is clear that both $T$ and $S$ are invertible. Hence the theorem
is true. \hfill$\Box$

\section{ Maps preserving  separable pure states and strict convex combinations: Multipartite systems}

The results similar to that in Section 3 for bipartite case are
valid for multipartite cases, of course, with   more complicated
expressions. The proofs are also similar. In this section we only
list some of them, which have relatively simple expressions and may
have more applications. The meanings  of the notations used here are
also similar to that in Section 3.

Suppose $\dim H_i=n_i$. For $1\leq r_1<\cdots <r_p \leq n$, define
the partial trace which is a linear map ${\rm
Tr}^{r_1,\cdots,r_p}:{\mathcal B}_{\rm
sa}(\otimes_{i=1}^nH_i)\rightarrow {\mathcal B}_{\rm sa}(\otimes
_{j=1}^pH_{r_j}) $ as follows: $$  \otimes
_{i=1}^nA_i\longmapsto(\prod_{i\neq{r_1,\cdots,r_p}}{\rm
Tr}A_i)\otimes _{j=1}^pA_{r_j}.$$ In particular, the linear map
${\rm Tr}^r:{\mathcal B}_{\rm sa}(\otimes_{i=1}^nH_i)\rightarrow
{\mathcal B}_{\rm sa}(H_r)$ is given by ${\rm
Tr}^r(\otimes_{i=1}^nA_i)=(\prod_{i\neq r}{\rm Tr}(A_i))A_r$. We
call ${\rm Tr}^r(\rho)$  the reduction state of $\rho\in{\mathcal
S}(H_1\otimes H_2\otimes\cdots\otimes H_n)$ in the subsystem
${\mathcal S}(H_r)$. A multipartite state $\rho\in{\mathcal
S}(\otimes_{i=1}^nH_i)$ is called a product state if
$\rho=\otimes_{i=1}^n\rho_i$ for some $\rho_i\in{\mathcal S}(H_i)$.

The following result corresponds to Corollary 3.3 and Corollary 3.4.

\textbf{Theorem 4.1} {\it Let $\psi:{\mathcal S}_{\rm
sep}(H_1\otimes H_2\otimes \cdots\otimes H_n)\rightarrow {\mathcal
S}_{\rm sep}(K_1\otimes K_2\otimes \cdots\otimes K_n)$ be a strict
convex combination preserving  map, with $2\leq \dim H_i<\infty$,
$i=1, 2, \cdots, n$, and
  $\psi(\mathcal
{P}ur(H_1)\otimes \mathcal {P}ur(H_2)\otimes \cdots \otimes \mathcal
{P}ur(H_n))\subseteq \mathcal {P}ur(K_1)\otimes \mathcal
{P}ur(K_2)\otimes \cdots \otimes \mathcal {P}ur(K_n)$.  }

(1) {\it If the range of $\psi$ is non-collinear or a singleton,
then $\psi$ maps product states to product states.}

(2) {\it If  the range of $\psi$ is non-collinear and contains a
state $\sigma$ so that its reduction state ${\rm Tr}^i(\sigma)$ has
rank $\geq 2$ for each $i=1,2,\ldots, n$, then there exist a
permutation $\pi:(1, \cdots, n)\mapsto (\pi(1), \cdots, \pi(n))$ of
$(1, \cdots, n)$ and injective linear or conjugate linear (may not
simultaneously)  operators $M_j:H_{\pi(j)}\rightarrow K_j$, $j=1,
\cdots, n$, such that
$$\psi(\rho)=\frac{(M_1\otimes \cdots\otimes
M_n)\Theta_\pi(\rho)(M_1^*\otimes \cdots\otimes M_n^*)}{{\rm
Tr}((M_1\otimes \cdots\otimes M_n)\Theta_\pi(\rho)(M_1^*\otimes
\cdots\otimes M_n^*))}$$ for all $\rho\in {\mathcal S}_{\rm
sep}(H_1\otimes \cdots \otimes H_n)$. Here $\Theta_\pi:{\mathcal
B}_{\rm sa}(H_1\otimes H_2\otimes \cdots\otimes H_n)\rightarrow
{\mathcal B}_{\rm sa}(H_{\pi(1)}\otimes H_{\pi(2)}\otimes
\cdots\otimes H_{\pi(n)})$ is a linear map determined by
$\Theta_\pi(A_1\otimes A_2\otimes \cdots\otimes
A_n)=A_{\pi(1)}\otimes A_{\pi(2)}\otimes \cdots \otimes A_{\pi(n)}$.
It is clear that $\dim H_{\pi(j)}\leq\dim K_j$.}

\if {\bf Proof.} We give  a skeleton of the proof. If the range of
$\psi$ is collinear, it is clear that either $\psi$ contracts to a
pure state or there exist distinct separable pure state $\sigma_1$
and $\sigma_2$ such that $\psi({\mathcal Pur}(H_1)\otimes {\mathcal
Pur}(H_2)\otimes\cdots\otimes {\mathcal Pur}(H_n))=\{\sigma_1,
\sigma_2\}$ and ${\rm ran}(\psi)\subseteq [\sigma_1, \sigma_2]$.
Thus, in the sequel, we assume the range of $\psi$ is non-collinear.
For $r=1,\cdots,n$, define maps $\phi_r:({\mathcal
S}(H_1),\cdots,{\mathcal S}(H_n))\rightarrow {\mathcal S}(K_r)$ by
$$\phi_r(A_1,\cdots,A_n)={\rm Tr}^r(\psi(\otimes_{i=1}^nA_i)) \ for \ all
\ (A_1,\cdots,A_n)\in ({\mathcal S}(H_1),\cdots,{\mathcal
S}(H_n)).$$ Notice that
$$\psi(\otimes _{i=1}^nP_i)=\otimes
_{r=1}^n\phi_r(P_1,\cdots,P_n) \ for \ all \ (P_1,\cdots,P_n)\in
(\mathcal {P}ur(H_1),\cdots,\mathcal {P}ur(H_n)).$$ Given arbitrary
$Q_i\in \mathcal {P}ur(H_i)$ for $i=2,\cdots,n$, the map
$\phi_r(\cdot,Q_2,\cdots,Q_n)$ maps $\mathcal {P}ur(H_1)$ into
$\mathcal {P}ur(K_r)$ and also   preserves strict convex
combination. Then by Theorem 2.5, the map
$\phi_r(\cdot,Q_2,\cdots,Q_n)$ must have the form (1) or (2) or (3)
stated in Theorem 2.5.

Firstly we claim that

{\bf Claim 1.} $\phi_r(\cdot,Q_2,\cdots,Q_n)$, $r=1,\cdots,n$, can
not have the form Theorem 2.5.(2) for any choice $(Q_2,\ldots, Q_n)$
with $Q_i\in \mathcal {P}ur(H_i)$.

As $\psi$ is strict convex combination preserving and ${\rm
ran}\psi$ is non-collinear,  by Lemma 2.3, we have
$\psi(\rho)=\frac{\Gamma(\rho)+B}{f(\rho)+b}$ for all $\rho\in
{\mathcal S}_{\rm sep}(H_1\otimes \cdots \otimes H_n)$, where
$\Gamma:\mathcal {B}_{\rm sa}(H_1\otimes \cdots \otimes
H_n)\rightarrow \mathcal {B}_{\rm sa}(H_1\otimes \cdots \otimes
H_n)$ is a linear map, $B\in \mathcal {B}_{\rm sa}(H_1\otimes \cdots
\otimes H_n)$ is an operator, $f:\mathcal {B}_{\rm sa}(H_1\otimes
\cdots \otimes H_n)\rightarrow \mathbb{R}$ is a linear functional
and $b\in \mathbb{R}$ with $f(\rho)+b>0$ for any $\rho\in {\mathcal
S}_{\rm sep}(H_1\otimes \cdots \otimes H_n)$. Since $H_1\otimes
H_2\otimes\cdots\otimes H_n$ is finite dimensional, $\Gamma$, $f$
are continuous. It follows that $\psi$ is continuous and so
$\phi_r(\cdot,Q_2,\cdots,Q_n)$ is continuous on ${\mathcal S}(H_1)$
for each $r=1,\cdots,n$, this ensures  that
$\phi_r(\cdot,Q_2,\cdots,Q_n)$ can not be the form Theorem 2.5.(2).

{\bf Claim 2.}
 For any $r\in\{1,2,\ldots,n\}$, either
$\phi_r(\cdot,Q_2,\cdots,Q_n)$ has the form Theorem 2.5.(1) for all
$Q_i\in \mathcal {P}ur(H_i)$ or $\phi_r(\cdot,Q_2,\cdots,Q_n)$ has
the form Theorem 2.5.(3) for all $Q_i\in \mathcal {P}ur(H_i)$.

Fix $A_1\in {\mathcal S}(H_1)$ with ${\rm rank}A_1\geq 2$. Define
$F_{A_1,r}:(\mathcal {P}ur(H_2),\cdots,\mathcal
{P}ur(H_n))\rightarrow \mathbb{R}$ by
$$F_{A_1,r}(Q_2,\cdots,Q_n)=\|\phi_r(A_1,Q_2,\cdots,Q_n)\|_2.$$ It is clear that, when
$\phi_{r}(\cdot,Q_2,\cdots,Q_n)$ has the form Theorem 2.5.(1),
$F_{A_1,r}(Q_2,\cdots,Q_n)=1$; when $\phi_r(\cdot,Q_2,\cdots,Q_n)$
has the form Theorem 2.5.(3), $F_{A_1,r}(Q_2,\cdots,Q_n)<1$. Then
similar to the argument of Claim 1 in the proof of Theorem 3.2, we
obtain that either $\phi_r(\cdot,Q_2,\cdots,Q_n)$ has the form
Theorem 2.5.(1) for all $Q_i\in \mathcal {P}ur(H_i)$ or
$\phi_r(\cdot,Q_2,\cdots,Q_n)$ has the form Theorem 2.5.(3) for all
$Q_i\in \mathcal {P}ur(H_i)$.

{\bf Claim 3.} For  any choice $(Q_2,\ldots,Q_n)\in (\mathcal
{P}ur(H_2),\ldots,{\mathcal Pur}(H_n))$, at most one of the maps
$\phi_r(\cdot,Q_2,\cdots,Q_n)$, $r=1,\cdots,n$, has the form Theorem
2.5.(3) and all the exceptional maps have the form Theorem 2.5.(1).

Assume that there exists $(Q_2,\ldots,Q_n)$ so that both the maps
$$\phi_s(\cdot,Q_2,\cdots,Q_n)\ {\rm and}\
\phi_t(\cdot,Q_2,\cdots,Q_n)$$
 have the form Theorem 2.5.(3) for some distinct $s,t$. Then consider the map $L:{\mathcal S}(H_1)\rightarrow
{\mathcal S}(K_s \otimes  K_t)$ defined by $L(A)={\rm
Tr}^{s,t}(\psi(A\otimes (\otimes_{i=2}^nQ_i)))$. Obviously, $L$ is
strict convex combination preserving and
$L(P)=\phi_s(P,Q_2,\cdots,Q_n)\otimes \phi_t(P,Q_2,\cdots,Q_n)\in
{\mathcal Pur}(K_s\otimes K_t)$ for all $P\in \mathcal {P}ur(H_1)$.
Because $\phi_s(\cdot,Q_2,\cdots,Q_n)$ and
$\phi_t(\cdot,Q_2,\cdots,Q_n)$ are of the form Theorem 2.5.(3),
there exist injective linear or conjugate linear operators $M_s$ and
$M_t$ such that $\phi_s(P,Q_2,\cdots,Q_n)=\frac{M_sPM_s^*}{{\rm
Tr}(M_sPM_s^*)}$ and $\phi_t(P,Q_2,\cdots,Q_n)=\frac{M_tPM_t^*}{{\rm
Tr}(M_tPM_t^*)}$, respectively. It follows that
$$L(P)=\frac{(M_s\otimes M_t)(P\otimes P)(M_s\otimes
M_t)^*}{{\rm Tr}((M_s\otimes M_t)(P\otimes P)(M_s\otimes M_t)^*)}.$$
Following the same argument as  Claim 2 in the proof of Theorem 3.2,
one can get a contradiction.

Applying the same argument on the map
$\phi_r(Q_1,\cdots,Q_{p-1},(\cdot),Q_{p+1},\cdots,Q_n):{\mathcal
S}(H_p)\rightarrow{\mathcal S}(K_r)$, one sees that Claim 1 to Claim
3 also hold for  any $p=2,\cdots,n$.

Now, by a similar approach as that in the proof of Theorem 3.2, it
is easily seen that one of the following holds: (i) $\psi $
contracts to a separable pure state; (ii) $\psi({\mathcal
Pur}(H_1)\otimes{\mathcal Pur}(H_2)\otimes\cdots{\mathcal
Pur}(H_n))$ contains two separable pure states $\sigma 1,\sigma 2$
and ran$(\psi)\subseteq [\sigma_1,\sigma_2]$; (iii) ran$(\psi)$ is
non-collinear,  $\psi$ maps product states to product states and,
for any $\rho\in{\mathcal S}_{\rm sep}(H_1\otimes
H_2\otimes\cdots\otimes H_n)$, at least one of reductions ${\rm
Tr}^i(\psi(\rho))$, $i=1,2,\ldots,n$, is a pure state; (iv) $\psi$
sends product states to product states and there is a permutation
$(\pi(1), \cdots, \pi(n))$ of $(1,\cdots, n)$ such that
$\phi_{\pi^{-1}(p)}(Q_1, \cdots, Q_{p-1},(\cdot), Q_{p+1},\cdots,
Q_n)$ has the form Theorem 2.5.(3) for all $Q_i\in \mathcal
{P}ur(H_i)$.

Therefore, if the range of $\psi$ is non-collinear or singleton,
then $\psi$ sends product states to product states, that is, the
statement (1) holds.

 Furthermore, if there exists $\rho\in{\mathcal S}_{\rm sep}(H_1\otimes
H_2\otimes\cdots\otimes H_n)$ such that all reductions ${\rm
Tr}^i(\psi(\rho))$, $i=1,2,\ldots,n$, have rank $\geq 2$,  then only
(iv) can hold. Thus, there exist a permutation $\pi$ of
$(1,2,\ldots, n)$ and injective linear or conjugate linear (may not
simultaneously) operators $M_i :H_{\pi(i)}\rightarrow K_i$ such that
$\phi_i(Q_1, \cdots, P_{\pi(i)},\cdots, Q_n) :{\mathcal
S}(H_{\pi(i)})\rightarrow{\mathcal S}(K_i)$ is of the form
$\rho_{\pi(i)}\mapsto\frac{M_i\rho_{\pi(i)}M_i^*}{{\rm
Tr}(M_i\rho_{\pi(i)}M_i^*)}$. Moreover,
$$\begin{array}{rl}& \psi(P_1\otimes \cdots \otimes P_n)=\phi_1(P_1,
\cdots, P_n)\otimes \cdots\otimes \phi_n(P_1,\cdots, P_n)\\=
&\phi_1(Q_1, \cdots, P_{\pi(1)}, \cdots, Q_n)\otimes \cdots\otimes
\phi_n(Q_1, \cdots, P_{\pi(n)},\cdots,
Q_n)\\=&\frac{M_1P_{\pi(1)}M_1^*}{{\rm
Tr}(M_1P_{\pi(1)}M_1^*)}\otimes \cdots\otimes
\frac{M_nP_{\pi(n)}M_n^*}{{\rm Tr}(M_nP_{\pi(n)}M_n^*)}\end{array}$$
for any $P_1\otimes \cdots \otimes P_n\in \mathcal {P}ur(H_1)\otimes
\cdots \otimes \mathcal {P}ur(H_n)$.  Then it is easily checked that
$$\psi(A_1\otimes \cdots A_n)=\frac{M_1A_{\pi(1)}M_1^*}{{\rm
Tr}(M_1A_{\pi(1)}M_1^*)}\otimes \cdots\otimes
\frac{M_nA_{\pi(n)}M_n^*}{{\rm Tr}(M_nA_{\pi(n)}M_n^*)}$$ for any
$A_1\otimes \cdots\otimes A_n\in {\mathcal S}_{\rm sep}(H_1\otimes
\cdots\otimes H_n)$. Now an argument similar  to the proof of
Corollary 3.4 entails that   $$\psi(\rho)=\frac{(M_1\otimes \cdots
\otimes M_n) \Theta_{\pi}(\rho) (M_1\otimes \cdots \otimes
M_n)^*}{{\rm Tr}((M_1\otimes \cdots \otimes M_n) \Theta_{\pi}(\rho)
(M_1\otimes \cdots \otimes M_n)^*)}$$ holds for all
$\rho\in{\mathcal S}_{\rm sep}(H_1\otimes H_2\otimes\cdots\otimes
H_n)$. Here  $\Theta_{\pi}:{\mathcal B}_{\rm sa}(H_1\otimes
H_2\otimes \cdots\otimes H_n)\rightarrow {\mathcal B}_{\rm
sa}(H_{\pi(1)}\otimes H_{\pi(2)}\otimes \cdots\otimes H_{\pi(n)})$
is the linear map determined by $\Theta_\pi(A_1\otimes A_2\otimes
\cdots\otimes A_n)=A_{\pi(1)}\otimes A_{\pi(2)}\otimes \cdots
\otimes A_{\pi(n)}$.   Obviously $\dim H_{\pi(i)}\leq \dim K_i$,
$i=1, \cdots, n$. So the statement (2) is true. \hfill$\Box$ \fi

The following result is corresponding to Theorem 3.5.

\textbf{Corollary 4.2} {\it Let $\psi:{\mathcal S}_{\rm
sep}(H_1\otimes H_2\otimes \cdots\otimes H_n)\rightarrow {\mathcal
S}_{\rm sep}(K_1\otimes K_2\otimes \cdots\otimes K_n)$ be a
bijective map with $2\leq \dim H_i<\infty$, $i=1, 2, \cdots, n$.
Then $\psi$ is strict convex combination preserving  if and only if
there exist a permutation $\pi$ of $(1, 2,\cdots, n)$  and
invertible linear or conjugate linear   (may not simultaneously)
operators $M_j:H_{\pi(j)}\rightarrow K_j$, $j=1, \cdots, n$, such
that
$$\psi(\rho)=\frac{(M_1\otimes \cdots\otimes
M_n)\Theta_\pi(\rho)(M_1^*\otimes \cdots\otimes M_n^*)}{{\rm
Tr}((M_1\otimes \cdots\otimes M_n)\Theta_\pi(\rho)(M_1^*\otimes
\cdots\otimes M_n^*))} \eqno(4.1)$$ holds for all $\rho\in {\mathcal
S}_{\rm sep}(H_1\otimes \cdots \otimes H_n)$. Here
$\Theta_\pi:{\mathcal B}_{\rm sa}(H_1\otimes H_2\otimes
\cdots\otimes H_n)\rightarrow {\mathcal B}_{\rm
sa}(H_{\pi(1)}\otimes H_{\pi(2)}\otimes \cdots\otimes H_{\pi(n)})$
is the linear map determined by $\Theta_\pi(A_1\otimes A_2\otimes
\cdots\otimes A_n)=A_{\pi(1)}\otimes A_{\pi(2)}\otimes \cdots
\otimes A_{\pi(n)}$. It is clear that $\dim H_{\pi(j)}=\dim K_j$.}

\if {\bf Proof.} For ``only if" part, it is clear that the
assumptions on $\psi$ implies that $\psi(\mathcal {P}ur(H_1)\otimes
\mathcal {P}ur(H_2)\otimes \cdots \otimes \mathcal {P}ur(H_n))=
\mathcal {P}ur(K_1)\otimes \mathcal {P}ur(K_2)\otimes \cdots \otimes
\mathcal {P}ur(K_n)$. Then by Theorem 4.1, $\psi$   has the form
Eq.(4.1) with each $M_j$ bijective. The ``if" part is obvious.
\hfill$\Box$\fi

\section{Conclusion}

  The quantum measurement  map  preserve strict convex combinations  and sends pure states
  to pure states. Similarly, each local
quantum measurement map   preserves separable pure states and strict
convex combinations. These facts  make it interesting to study the
problem of characterizing the maps between   states in
(multipartite) quantum systems that are (separable) pure state
preserving and strict convex combination preserving. These problems
are basic and interesting both in quantum information science and
mathematics science, and their solutions will present a geometric
characterization of (local) quantum measurements and help us to
understand better the quantum measurement.

In the present paper, we  give a characterization of the maps
$\phi:{\mathcal S}(H)\rightarrow{\mathcal S}(K)$ with $2\leq \dim
H<\infty$ that preserve pure states and strict convex combinations,
and give a structure theorem of the maps $\psi:{\mathcal
S}(H_1\otimes H_2\otimes \cdots\otimes H_n)\rightarrow{\mathcal
S}(K_1\otimes K_2\otimes\cdots\otimes K_n)$ with $2\leq \dim
H_i<\infty$ that preserve separable pure states and strict convex
combinations.    From these results we get a characterization of
injective (local) quantum measurements. In almost all situations,
for example, in the case that the range of $\psi$ is non-collinear
or a singleton,   $\psi$ sends product states to product states. In
particular, if   the range of $\psi$ is non-collinear and contains
an element with each reduction having rank $\geq 2$, then $\phi$ is
essentially an injective local quantum measurement. Thus we answer
affirmatively two conjectures proposed in \cite{HL}.

Finally we remark that, the main results obtained in \cite{HHL,HL}
hold for both finite dimensional systems and infinite dimensional
systems. However, in the present paper we only deal with  finite
dimensional systems. We conjecture that the results in this paper
are also valid for infinite dimensional case but new tools are
needed to prove this.

{\bf Acknowledgement.} The authors wish to give their thanks to the
referees for helpful comments and suggestions to improve the
expression of this paper.


\begin{thebibliography}{99}

\bibitem{AS}  Alfsen, E., Shultz, F., Unique decompositions, faces, and
automorphisms of separable states, Journal of Mathematical Physics
51(2010), 052201.


\bibitem{BZ} Bengtsson, I.,  Zyczkowski, K.,
Geometry of Quantum States, An introduction to quantum entangument,
Cambridge University Press, Cambridge, 2006.

\bibitem{FLPS} Friedland, S., Li C.-K., Poon Y.-T. and Sze N.-S. The
automorphisms group of separable states in quantum information
theory  J. Math. Phys. 52 (2011) 042203.

\bibitem{G} Gudder, S.,  A structure for quantum measurements, Reports on Mathematical Physics, 55 (2005) 2, 249-267.


\bibitem{HHL}   He, K.  Hou, J. and Li C.-K., A geometric characteristic of  invertible quantum measurement maps, J. Funct. Anal. 264 (2013), 464-478.

\bibitem{H1}  Hou, J., A characterization of
positive linear maps and  criteria for entangled quantum states, J.
Phys. A: Math. Theor. 43 (2010) 385201.


\bibitem{H3} Hou, J., On the tensor products of operators, Acta Math. Sinica
(New Ser.), 9 (1993), 195-202.

\bibitem{HL} Hou, J.   and Liu, L., Quantum measurement and maps
preserving convex combinations of separable states, J. Phys. A:
Math. Theor. 45 (2012) 205305.

\bibitem{HQ} Hou J. and Qi X., Linear maps preserving separability of pure
states,  arXiv:1210.2155

\bibitem{MN} Mendonca, P. E. M. F., Napolitano, R. d. J., etc.,
Alternative fidelity measure between quantum states, Phy. Rev. A 78
(2008) 052330.










\bibitem{NC} Nielsen, M.  A.,  Chuang, I. L., Quantum Computation and
Quantum Information, Cambridge University Press, Cambridge, 2000.

\bibitem{Z} Z. P$\breve{a}$les,  Characterization of segment and convexity preserving maps,
arXiv:1212.1268v1.



\bibitem{VPJK} Vedral, V., Plenio, M. B.,  Jacobs, K., Knight, P. L.,  Phys.
Rev. Lett. 78 (1997) 2275; Phys. Rev. Lett. 57 (1997) 4452

\bibitem{W} Werner, R. F.,
Quantum states with Einstein-Podolsky-Rosen correlations admitting a
hidden-variable model, Phys. Rev. A 40 (1989) 4277.


\end{thebibliography}
\end{document}